\documentclass[twocolumn]{aastex62}

\usepackage{lineno}
\defcitealias{K15}{K15}
\defcitealias{Guo2013PhotoCat}{G13}

\begin{document}

\title{\MakeUppercase{Strategies for obtaining robust SED fitting parameters for galaxies at \lowercase{z}$\sim$1 and \lowercase{z}$\sim$2 in the absence of IR data}}

\author{Chandler Osborne}
\affil{Department of Astronomy, Indiana University, Bloomington, IN, 47408}

\author{Samir Salim}
\affil{Department of Astronomy, Indiana University, Bloomington, IN, 47408}

\begin{abstract}

Robust estimation of star formation rates (SFRs) at higher redshifts ($z \gtrsim 1$) using UV-optical-NIR photometry is contingent on the ability of spectral energy distribution (SED) fitting to simultaneously constrain the dust attenuation, stellar metallicity, and star formation history (SFH).   IR-derived dust luminosities can help break the degeneracy between these parameters, but IR data is often not available. Here, we explore strategies for SED fitting at $z \gtrsim 1$ in the absence of IR data using a sample of log $M_{*} > 10.2$ star-forming galaxies from the Cosmic Assembly Near-infrared Deep Extragalactic Legacy Survey (CANDELS) for which 24$\mu$m data are available. We adopt the total IR luminosity ($L_{\text{TIR}}$) obtained from 24$\mu$m as the ``ground truth'' that allows us to assess how well it can be recovered (as $L_{\text{dust}}$) from UV-optical-NIR SED fitting. We test a variety of dust attenuation models, stellar population synthesis models, metallicity assumptions, and SFHs separately to identify which assumptions maximize the agreement (correlation and linearity) between $L_{\text{TIR}}$ and $L_{\text{dust}}$.  We find that a flexible dust attenuation law performs best.  For stellar populations, we find that \citet{BC03} models are favored over those of \citet{Eldridge&Stanway+2017BPASS2pt1}.  Fixing the stellar metallicity at solar value is preferred to other fixed values or leaving it as a free parameter.  For SFHs, we find that minimizing the variability in the recent ($<$ 100 Myr) SFH improves the agreement with $L_{\text{TIR}}$. Finally, we provide a catalog of galaxy parameters (including $M_*$and SFR) for CANDELS galaxies with $\log M_*>8$ and $0.7<z<1.3$ obtained using the models we found to be the most robust.

\end{abstract}

\section{Introduction}

The current star formation rate (SFR) and stellar mass are two of the most fundamental properties of galaxies.  For an accurate understanding of galaxy formation and evolution, robust constraints of both properties for statistically significant samples are needed.  Galaxy properties are often estimated by fitting libraries of model spectra to observed spectral energy distributions (SEDs), a process referred to as SED fitting \citep[for a review, see][]{Walcher2011SEDfittingReview,Conroy2013SEDFittingReview}.  In recent years, the practice of SED fitting has been greatly refined.  State of the art SED fitting codes such as CIGALE \citep{BoquienCIGALE2019}, PROSPECTOR \citep{Leja2017ProspectorSEDCode, Johnson2021ProspectorSEDFitting}, MIRKWOOD \citep{Mirkwood2021SEDfittingCode}, BAGPIPES \citep{Carnall2018BagpipesSEDcode}, BEAGLE \citep{Chevallard&Charlot2016BeagleSEDcode}, and SEDfit \citep{Sawicki2012SEDfitCodeOG} are highly flexible, efficient, and accessible, and offer a diverse set of models to choose from.

In modeling galaxy SEDs, there are three crucial components: a stellar population synthesis (SPS) model, a star formation history (SFH), and a dust attenuation model.  The adopted SPS model is convolved with a star formation history, which specifies the star formation rate as a function of time, to produce a model spectrum of a composite stellar population with stars of various ages.  The spectrum is then attenuated according to some dust law(s). Model SEDs with stellar populations of various ages and metallicities and attenuated by varying levels of dust are then fit to observed galaxy SEDs or spectra.

Age, metallicity, and dust each act to redden the spectrum \citep{BelldeJong2001StellarMtoLandTFrelation, Carter2009ageMetalDegenforSPS}.  As the signal-to-noise of the photometry degrades, the age-dust-metallicity degeneracy is exacerbated and the accuracy and/or precision of the SED fitting may be compromised. Progress has been made in identifying sets of models which are physically motivated and help to minimize the effects of degeneracy in the fitting \citep[e.g.,][]{Pacifici2012relativeMeritsOfDiffObservationsForSEDmodeling}.  It is well-established that galaxies host a wide variety of dust attenuation laws with different UV bump strengths and steepness, and allowing for this flexibility in the SED fitting produces more accurate results, at least at low redshifts \citep{Salim2018DustAttCurves}.  SPS models are now available that include the effects of binary stars, which may be needed to infer accurate physical properties especially for young stellar populations \citep{Eldridge&Stanway+2017BPASS2pt1, Stanway&Eldridge2018BPASSoldpops}.  Well-motivated SFH models have proven effective in correcting for bias in stellar mass estimates due to outshining \citep{Buat2014reliabilityOfSEDfittingAtzgtr1, Michalowski2014massesOfSubmmGalaxiesFavor2cSFHs, Simha2014ParameterizingSFHs, Sorba2015outshining, Salim2016GSWLC, Lower2020StellarMassSEDFittingSFHBias}, and can produce self-consistent measurements of the stellar mass growth and star formation rate density over time \citep{Leja2019OlderMoreQuiescentUniverse, Leja2020resolvedSMF&SFRDdiscrepancy}.  Recently, state-of-the-art SFHs have been utilized to reconcile the long-standing discrepancy between the normalization of the star-forming main sequence from simulations with that inferred from observations \citep{Leja2022resolvedSFMSdiscrepancyObs&Sims}.  The SED fitting community continues to make great strides in improving the accuracy and reliability of galaxy spectral modelling.

Improvements in accuracy are not necessarily accompanied by improvements in precision, however, and model degeneracies will tend to inflate the uncertainties of physical parameters even when systematic offsets are well-accounted for.  Complementing UV and optical photometry with mid- and far-IR photometry, which is sensitive to the emission of dust heated by young stars, can be used to break the age-dust-metallicity degeneracy and greatly improve the constraining power of the SED fitting \citep[e.g.,][]{daCunha2008SimpleModelForUVtoIRGalaxyEmission, Buat2014reliabilityOfSEDfittingAtzgtr1}.  Unfortunately, however, IR surveys with Spitzer and Herschel were limited in their completeness and sensitivity even for massive galaxies, and for most galaxies one is limited to fitting only their UV-optical-NIR photometry where age, dust, and metallicity must be simultaneously constrained.  It is therefore crucial to understand which model choices ensure that the SED fitting is as precise as possible.  The need for good general SED fitting practices is especially important with the imminent influx of survey data from JWST as well as the Vera C. Rubin Observatory.

In this work, we identify some good practices for obtaining robust physical properties from SED fitting of UV-optical-NIR photometry without constraints from the mid- or far-IR for galaxies at $z \sim 1$ and $z \sim 2$.  Our sample consists of star-forming galaxies with well-sampled UV-optical-NIR SEDs and solid mid-IR photometry.  We explore the effects of different models for dust, stellar metallicity, SPS model, and star formation histories on the reliability of physical parameters inferred from the SED fitting.  In particular, we determine a set of models for which the total energy absorbed by dust in the UV-optical-NIR ($L_{\text{dust}}$) best agrees (in terms of correlation and linearity) with the total IR luminosity ($L_{\text{TIR}}$) inferred from mid-IR photometry.

% Go over paper sections here and finish intro

This paper is organized as follows.  We describe our data and sample selection in Section \ref{Section::Data&SS}.  We detail our analysis methods, including choice of SED fitting code and choice of models, in Section \ref{Section:Methods}.  We then present our results in Section \ref{Section:Results}, discuss these results in Section \ref{Section:Discussion}, and conclude in Section \ref{Section:Conclusions}. We assume a flat WMAP7 cosmology ($H_0 = 70$ km/s/Mpc, $\Omega_m = 0.27$) throughout the paper.

\section{Data $\&$ Sample Selection}
\label{Section::Data&SS}

\begin{figure}
    \centering
    \includegraphics[scale=0.7]{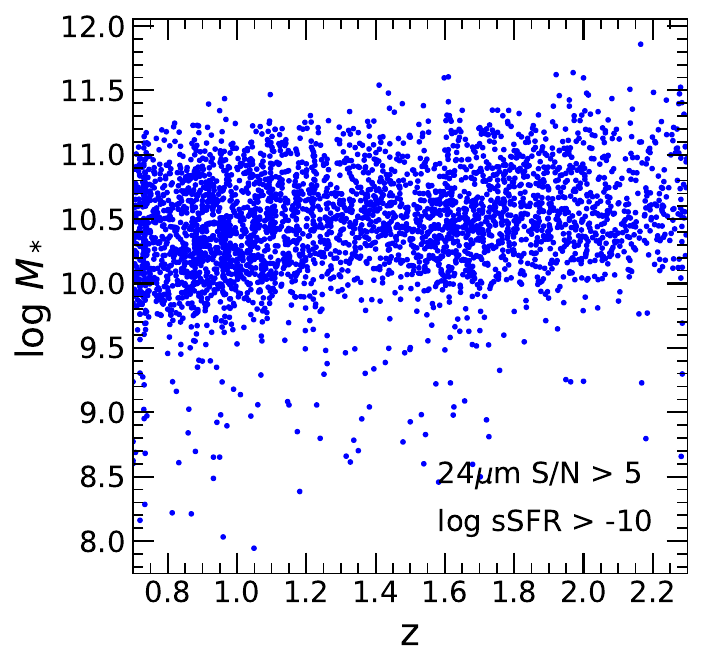}
    \caption{Stellar mass versus redshift distribution for star-forming galaxies (log sSFR $> -10$) in our sample with robust IR detections ($24\mu$m S/N $> 5$).  To ensure consistent completeness at all redshifts, we include only galaxies with log $M_* > 10.2$ in our final sample.}
    \label{Mstar_versus_z}
\end{figure}

\begin{figure*}
    \centering
    \gridline{\fig{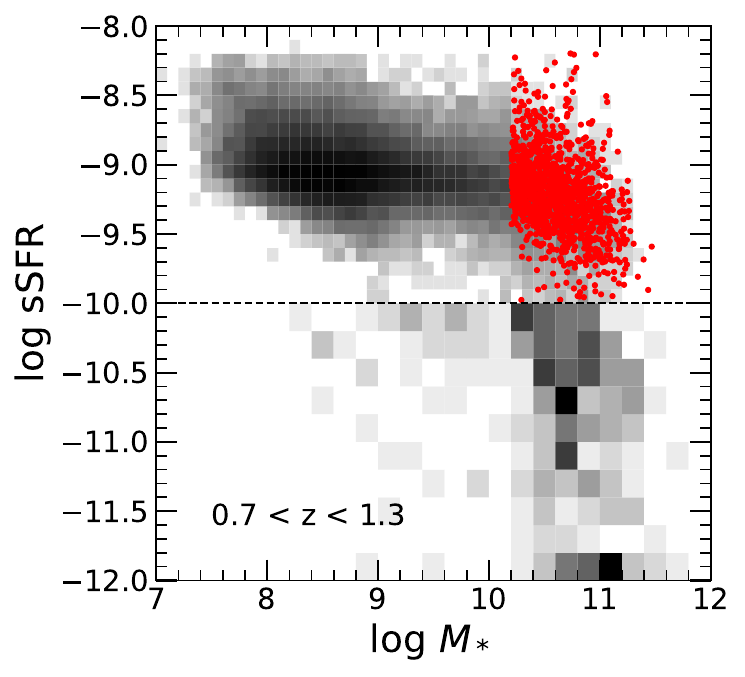}{0.45\textwidth}{}
          \fig{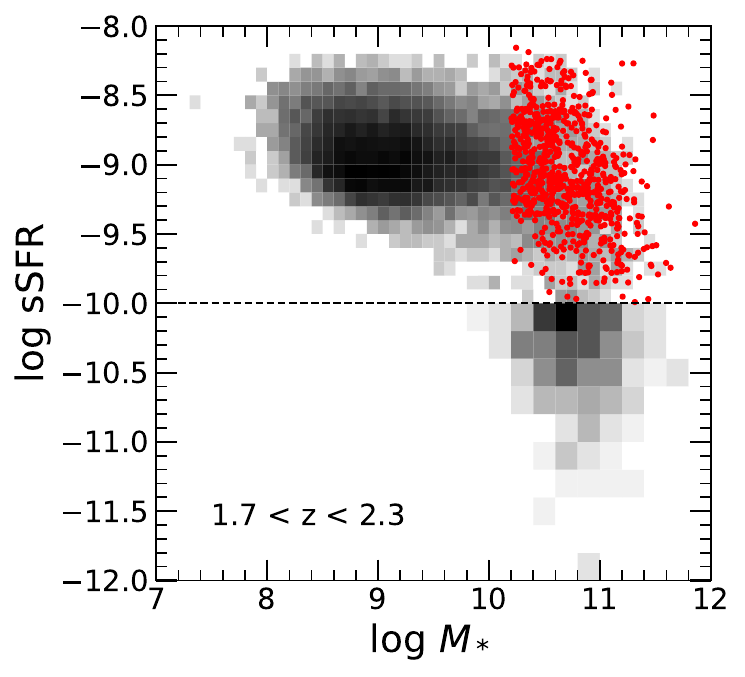}{0.45\textwidth}{}
          }
    \caption{Specific star formation rate versus stellar mass for galaxies in two redshift windows considered in this study.  We separate quiescent (plus quenching) and star forming galaxies using a log sSFR $= -10$ cut.  The entire sample of CANDELS galaxies on which we performed the SED fitting are shown as a gray density plot.  Our final sample, for which we require a robust IR detection ($24\mu$m signal to noise $> 5$), active star formation (log sSFR$ > -10$), and a relatively high mass (log $M_* > 10.2$), are shown as red points.  Current IR observations are not able to detect most of the lower-mass galaxies in HST images.}
    \label{ssfr_vs_mstar_sample}
\end{figure*}

Our sample of galaxies with photometry in the rest-frame UV-MIR regime is drawn from CANDELS.  We focus on four of the five CANDELS fields with the most comprehensive photometry: EGS, GOODS-S, UDS, and COSMOS.  We use publicly available photometric catalogs for each field: \citet{Stefanon2017EGSphotometry} for EGS, \citet{Guo2013PhotoCat} for GOODS-S, \citet{Galametz2013UDSphotometry} for UDS, and \citet{Nayyeri2017COSMOSphotometry} for COSMOS.  The instruments and passbands used for the SED fitting in this work are Blanco (\textit{U}), VLT/VIMOS (\textit{U}), HST/ACS (F435W, F606W, F755W, F814W, \textit{z}\textsubscript{goods}), HST/WFC3 (F098M, F105W, F125W, F140W, F160W), CFHT/MegaCam(\textit{u}, \textit{g}, \textit{r}, \textit{i}, \textit{z}), CFHT/WIRCAM (\textit{J}, \textit{H}, \textit{K}) Subaru/SuprimeCam (\textit{B}, \textit{g}, \textit{V}, \textit{R}\textsubscript{C}, \textit{r}, \textit{i}, \textit{z}), UKIRT/WFCAM (\textit{J}, \textit{H}, \textit{K}), Mayall/NEWFIRM (\textit{J}\textsubscript{1}, \textit{J}\textsubscript{2}, \textit{J}\textsubscript{3}, \textit{H}\textsubscript{1}, \textit{H}\textsubscript{2}, \textit{K}\textsubscript{s}), VLT/VISTA (\textit{Y}, \textit{J}, \textit{H}, \textit{K}\textsubscript{s}), VLT/ISAAC (\textit{K}\textsubscript{s}), VLT/HAWK-I (\textit{Y}, \textit{K}\textsubscript{s}), and VLT/HAWK-K (\textit{K}\textsubscript{s}).  Spitzer/IRAC photometry (Channels 1-4) are also utilized for each field from the GOODS \citep{GOODSsurvey2004IRSAref1, GOODSsurvey2011IRSAref2}, SEDS \citep{AshbySEDS2013IRACphotosurveyref}, SpUDS \citep{Dunlop2007surveySPUDSreference}, S-CANDELS \citep{Ashby2015_S-CANDELSsurveyRef}, S-COSMOS \citep{Sanders2007_S-COSMOS_SpitzerIRACsurveyref}, and AEGIS \citep{Davis2007AEGISsurveyRef, Barmby2008catalogOfMidIRsourcesInEGS} surveys.  These fields also have coverage at 24$\mu$m with Spitzer/MIPS \citep{GOODSsurvey2004IRSAref1, Davis2007AEGISsurveyRef, Dunlop2007surveySPUDSreference, Sanders2007_S-COSMOS_SpitzerIRACsurveyref} which we use to estimate total IR luminosities ($L_{\text{TIR}}$).  

To account for photometric zero-point errors, for each band except Spitzer/MIPS $24\mu$m (which we do not include in the SED fitting) we add to the reported error an amount equal to $10\%$ of the reported flux.  To be included in the SED fitting, we require a galaxy to have a detection in at least one band covering the rest-frame UV and be detected in at least 5 total bands in the rest-frame UV or rest-frame optical/near-IR range (again not counting Spitzer/MIPS $24\mu$m).  140,773 total galaxies across all four fields satisfy these photometric requirements.

\subsection{Redshift selection}

CANDELS provides photometric as well as spectroscopic redshifts.  We adopt the spectroscopic redshift where available and the photometric redshift otherwise; $\sim 10\%$ of galaxies have spectroscopic redshifts.  We impose a redshift cut of $z > 0.7$ to ensure that the photometry covers the rest-frame UV.  We furthermore exclude galaxies with $z > 2.3$ as the sample size diminishes significantly above this point.  In order to discern a potential redshift dependence in our results, we will in some instances divide our sample into two redshift windows, one at $0.7 < z < 1.3$ ($z \sim 1$) and the other at $1.7 < z < 2.3$ ($z \sim 2$); we otherwise include all galaxies in the range $0.7 < z < 2.3$.  We also require galaxies to have a CANDELS redshift that differs from the 3DHST \citep{Brammer20123DHSTprelimdetails, Skelton20143DHSTphotocatalog, Momcheva20163DHSTdatarelease} redshift by no more than 0.4 (i.e., $|\Delta z| < 0.4$) to avoid including galaxies with poor photometric redshifts.  Imposing the upper and lower redshift limits ($0.7 < z < 2.3$) reduces our sample to 76,593 galaxies, and the $|\Delta z| < 0.4$ requirement reduces it further to 60,266 galaxies.

\subsection{Final sample selection}

We further require that each galaxy have a 24$\mu$m detection with a signal-to-noise of at least 5, which reduces our sample size to 3,652 galaxies. We also impose a specific SFR cut (log sSFR $> -10$) so that we include only non-quenched/quenching star-forming galaxies in our assessments.  We note that there are very few $24\mu$m detections at log sSFR $< -10$ anyway.  We additionally impose a stellar mass cut (log $M_{*} > 10.2$) to ensure completeness at all the redshifts spanned by our sample (Figure \ref{Mstar_versus_z}). We find that our mass completeness limit is consistent with other estimates from the literature \citep[e.g.,][]{Chartab2020CandelsMassCompleteness}. Note that both the sSFR and the stellar mass used in these cuts are taken from the fiducial SED fits described in Section \ref{Section:FiducialModelPriors}.  We finally exclude galaxies with very poor fiducial SED fits (reduced $\chi^2 > 10$).  Applying the sSFR, stellar mass, and reduced $\chi^2$ cuts reduces our sample to 2,622 galaxies.  

Figure \ref{ssfr_vs_mstar_sample} shows the specific SFR versus stellar mass diagrams for the two redshift windows.  Shown in gray is our original CANDELS sample with photometric, redshift, $\Delta z$, and $\chi^2_r$ cuts applied, while the $24\mu$m-detected sample with the additional sSFR and stellar mass cuts (i.e., our final sample) applied is shown as red points.  The sample sizes are 1,221 at $z \sim 1$ and 757 at $z \sim 2$. 

\begin{figure}
    \centering
    \includegraphics[scale=0.7]{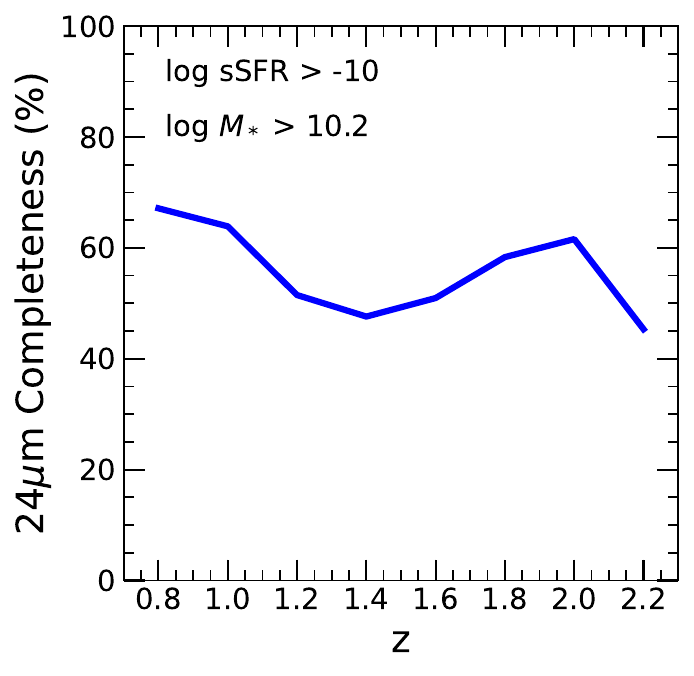}
    \caption{CANDELS Spitzer MIPS ($24\mu$m) completeness versus redshift for star-forming (log sSFR $> -10$), high-mass (log $M_* > 10.2$) galaxies in our sample.  Completeness is defined as the fraction of galaxies with S/N $> 5$ in the Spitzer $24\mu$m passband.}
    \label{IR_completeness_vs_z}
\end{figure}

Figure \ref{IR_completeness_vs_z} shows the completeness of the $24\mu$m photometry as a function of redshift for high-mass star-forming galaxies in our sample.  We define the completeness as the fraction of galaxies with $24\mu$m S/N $> 5$.  Notably, the $24\mu$m completeness decreases with redshift from $\sim 60\%$ at $z \sim 0.8$ to $\sim 40\%$ at $z \sim 1.5$, before rising back up to $\sim 60\%$ at $z \sim 2$.  This is due to a shifting of the rest-frame wavelength to $\sim 8\mu$m at $z \sim 2$ where strong PAH features are present in the IR spectra.  Overall, however, the completeness is relatively uniform across the entire redshift range.

\section{Methods}
\label{Section:Methods}

In this section we describe our choice of SED fitting code, our fiducial set of models which are used as a baseline, the various test models, our method for estimating the total IR luminosity ($L_{\text{TIR}}$) from the IR photometry, and our treatment of redshift-dependent systematic offsets between $L_{\text{TIR}}$ and $L_{\text{dust}}$.

\subsection{SED fitting}
\label{Section:SEDfittingwithCIGALE}

We use the Python-based SED modeling and fitting code CIGALE (version 2022.0) \citep{BoquienCIGALE2019}.  CIGALE allows specification of the star formation history (SFH), SPS model, nebular emission, and dust attenuation curve. CIGALE uses a Bayesian methodology wherein a probability distribution function (PDF) is constructed for each parameter (e.g., stellar mass) based on the model likelihoods.  The nominal value of the parameter is taken to be the mean of this PDF, and the standard deviation of the PDF gives the error \citep[see e.g.,][]{Smith2015DerivingSFHsEnergyBalanceMedianPDF}.  

CIGALE comes with a variety of built-in SED fitting `modules' which encompass the various models used in the SED fitting.  For example, two of the dust attenuation modules used in this work are `dustatt\textunderscore{}calzleit' and `dustatt\textunderscore{}powerlaw' which parameterize the dust attenuation curve in different ways.  We also use two modules that were created specifically for this work by modifying existing ones.  The first of these is the `dustatt\textunderscore{}calzleit\textunderscore{}2comp' module, which was created by modifying the built-in `dustatt\textunderscore{}calzleit' to allow the slope of the young and old attenuation curves to vary independently.  The second is the `bpass' module, which was created by modifying the built-in `bc03' module to use the Binary Population and Spectral Synthesis (BPASS, version 2.2) SPS model library \citep{Eldridge&Stanway+2017BPASS2pt1} instead of the \citet{BC03} (BC03) SPS model library.

\subsection{SED Fitting Models}
\label{Section:FiducialModelPriors}

\begin{deluxetable*}{ccccccc}[ht!]
\tablecaption{Dust attenuation law models.}
\tablecolumns{7}
\tablenum{1}
\label{DustRunsTable}
\tablewidth{0pt}
\tablehead{
\colhead{Model} &
\colhead{UV bump} & \colhead{Slope (young)} & \colhead{Slope (old)} & \colhead{$\frac{E(B-V)_{\text{old}} }{ E(B-V)_{\text{young}}}$} & \colhead{Normalization} & \colhead{CIGALE Module}
}
\startdata
Flexible (Fiducial) & 0 to 4.5 & -1.2 to 0.4 & Same as young & 0.44 & $E(B-V)_{\text{young}}$ = 0 to 1 & dustatt\textunderscore{}calzleit \\
Calzetti & 0 & 0 & Same as young & 1 & $E(B-V)_{\text{young}}$ = 0 to 1 & dustatt\textunderscore{}calzleit \\
Extra Flexible & 0 to 4.5 & -1.2 to 0.4 & -1.2 to 0.4 & 0.44 & $E(B-V)_{\text{young}}$ = 0 to 1 & dustatt\textunderscore{}calzleit\textunderscore{}2comp \\
Flexible (Power law) & 0 to 4.5 & 0.2 to 1.2 & Same as young & 0.44 & $A_{\text{V, young}}$ = 0 to 2.5 & dustatt\textunderscore{}powerlaw \\
\enddata
\tablecomments{Module `dustatt calzleit 2comp' is not part of a standard CIGALE distribution.}
\end{deluxetable*}

\begin{deluxetable*}{ccccc}[ht!]
\tablecaption{Stellar population models.}
\tablecolumns{5}
\tablenum{2}
\label{MetalRunsTable}
\tablewidth{0pt}
\tablehead{
\colhead{Model} & \colhead{SPS} & 
\colhead{$Z_*$} & \colhead{CIGALE Module}
}
\startdata
Solar $Z_*$ (Fiducial) & BC03 & 0.02 & bc03 \\
Free $Z_*$ & BC03 & 0.004, 0.008, 0.02, 0.05 & bc03 \\
Subsolar $Z_*$ & BC03 & 0.008 & bc03 \\
Solar $Z_*$ BPASS & BPASS & 0.02 & bpass \\
\enddata
\tablecomments{For all models we assume a \citet{ChabrierIMF2003} IMF with an upper mass limit of 100$M_{\odot}$. Module `bpass' is not part of a standard CIGALE distribution.}
\end{deluxetable*}

\begin{deluxetable*}{ccccccc}[ht!]
\tablecaption{Star formation history models.}
\tablecolumns{6}
\tablenum{3}
\label{SFHRunsTable}
\tablewidth{0pt}
\tablehead{
\colhead{Model} & 
\colhead{Character of old population} & \colhead{Burst Age} & \colhead{$\tau_{\text{burst}}$} & \colhead{CIGALE Module}
}
\startdata
Flat Burst (Fiducial) & Exponential & 100 - 1000 Myr & 20 Gyr  & sfh2exp \\
Flexible Burst & Exponential & 100 - 1000 Myr & 10 Myr - 20 Gyr & sfh2exp \\
Younger Flat Bursts & Exponential & 10 - 1000 Myr & 20 Gyr & sfh2exp \\
Delayed Old SFH & Delayed Exponential & 100 - 1000 Myr & 20 Gyr & sfhdelayed \\
\enddata
\tablecomments{For all models, the age of the old population is fixed at $\sim 500$ Myr less than the age of the universe at the time of observation and the burst fractions are limited to the range [0, 0.5].}
\end{deluxetable*}

In this section we describe the fiducial models used in the SED fitting, which we define as the models that produce the tightest agreement between $L_{\text{TIR}}$ and $L_{\text{dust}}$.  We also describe each variation on our fiducial model.  We independently vary the treatment for dust, metallicity, choice of SPS model, and SFH to isolate their effects on the recovered $L_{\text{dust}}$ from one another.  

All models include nebular emission (continuum and lines) for which we fix the gas-phase metallicity at $Z_* = 0.014$ and the ionization parameter at log $U = -3.4$.  While inclusion of some nebular emission is necessary to achieve unbiased results \citep{Pacifici2015importanceOfAppropriateSpectralModelsHighz, Salim2016GSWLC}, we find that varying the nebular parameters (e.g., using log $U = -2.8$) has negligible impact on $L_{\text{dust}}$ and thus SFR.

For all models explored in this work, we include all available photometry with the exception of IRAC channel 3 for all galaxies at $z < 1.1$ and IRAC channel 4 for all galaxies at $z < 1.5$ to exclude passbands that include hot dust emission.  

\subsubsection{Dust attenuation}
\label{Section:Variations:Dust}

A summary of the different dust models tested is shown in Table \ref{DustRunsTable}.  In our fiducial model the dust attenuation curve has a range of slopes and UV bump strengths, and is labelled as `flexible'. This flexible dust law follows the \citet{Noll2009CIGALE} parameterization as formulated by \citet{Salim2018DustAttCurves} and is based on the functional form of the \citet{Calzetti2000SFGDust} law, but has three important changes that provide it with greater flexibility: 1) it is a two-population law \citep[young and old population, see][]{Charlot&Fall2000DustCurves, Wuyts2009SimulatedGalaxiesLessonsForSEDModeling, Wuyts2011GalaxyStarFormationOF44} with higher attenuation around young stars ($\leq 10$ Myr, $E(B-V)_{\text{old}} / E(B-V)_{\text{young}} = 0.44$); 2) the law can have a range of steepness ($\delta$), i.e., the relative attenuation in the UV compared to the optical, which we allow to vary from significantly steeper than the Calzetti curve ($\delta = -1.5$) to shallower than the Calzetti curve ($\delta = 0.4$; note that the Calzetti curve has $\delta = 0$); and 3) a 2175\AA{} UV bump can be added, for which we allow the strength (i.e., the amplitude) to vary from 0 (no bump) to 4.5 (the bump in the Milky Way extinction curve has a strength of 3).  We allow the normalization $E(B-V)$ of the curve for the young population to vary from 0 to 1, with finer intervals at lower values.  

We test three dust models in addition to the fiducial model.  We first explore the simple assumption of a fixed Calzetti law \citep{Calzetti2000SFGDust}, which we refer to simply as the `Calzetti' model.  This model differs from fiducial by fixing the slope $\delta$ at zero and bump strength at zero (no bump).  The Calzetti model assumes equal attenuation for young and old stars ($E(B-V)_{\text{old}} / E(B-V)_{\text{young}} = 1$). 

The next model we consider is a modification of the fiducial dust law where we allow the slope $\delta$ of the young and old components to vary independently, which we refer to as the `extra flexible' model.  For this model the $\delta$ for both young and old curves have the same range of values as the fiducial model.  Like the fiducial model, the reddening of the old curve is normalized to 0.44 times the young curve.  

We also consider a flexible attenuation law which uses as its basis the power-law functional form, rather than the Calzetti one, which we label as `flexible (power law)'.  We allow the slope to vary from shallow ($n = 0.2$) to very steep ($n = 1.6$), again assuming the same slope for old and young components.  We also include a UV bump with the same range of amplitudes as the other models.  In CIGALE the power law curves are normalized in terms of $A_V$ instead of $E(B-V)$ and we allow this normalization to vary from 0 to 2.5 mag, covering a similar range of values in terms of $E(B-V)$ as the fiducial model.  Like the fiducial model, the reddening of the old curve is normalized to 0.44 times the young curve.  

\subsubsection{Stellar population synthesis models and metallicity}
\label{Section:Variations:StellarPops}

A summary of our stellar population model variations are shown in Table \ref{MetalRunsTable}.  Our SPS models are those of \citet{BC03}, hereafter referred to as BC03 models, with the stellar metallicity fixed at solar ($Z_* = 0.02$).  We label the fiducial model as `solar $Z_*$'.  A \citet{ChabrierIMF2003} IMF is assumed for all models.  

We test another three models in addition to the fiducial model. As an alternative to BC03, we consider the BPASS (version 2.2) stellar population models \citep{Eldridge&Stanway+2017BPASS2pt1, Stanway&Eldridge2018BPASSoldpops}.  BPASS is notable for including the effects of binary evolution which are expected to be significant for young, massive stars, and has been used extensively to study the ionizing continuum and stellar populations of galaxies at the dawn of reionization.  We compare the fiducial model to a BPASS model where the metallicity is also fixed at $Z_* = 0.02$, which we label as `solar $Z_*$ BPASS'. Using BPASS, we can assess the precision of $L_{\text{dust}}$ under different treatments of stellar evolution.  

We also test a variation of the fiducial model for which the metallicity is allowed to vary between $Z_* = 0.004$, $Z_* = 0.008$, $Z_* = 0.02$, and $Z_* = 0.05$, and label this model as `free $Z_*$'.  We finally test a variation of the fiducial model where the metallicity is fixed at subsolar ($Z_* = 0.008$), which we refer to as `subsolar $Z_*$'.

\subsubsection{Star formation histories}
\label{Section:Variations:SFH}

A summary of the SFH model variations used in this work are shown in Table \ref{SFHRunsTable}.  For the fiducial SFH we employ a parametric double exponential model which assumes an old population of fixed age with a second (burst) component superimposed.  The age of the old population is fixed and set to be $\sim 500$ Myr less than the age of the universe at the time of observation.  The e-folding time of the old population is allowed to vary from 200 Myr to 20 Gyr with a finer grid among shorter e-folding times in order to sample log sSFR space evenly.  We fix the e-folding time of the burst at 20 Gyr (functionally constant); thus our `burst' (in the fiducial model) is not transient but is simply another episode of SF.   We vary the fraction of mass formed in the burst from 0 to 0.5 and vary the age of the burst from 100 Myr to 1 Gyr, again with a finer grid at lower values.  Fixing the age of the old population and limiting the burst fraction and burst age prevents outshining of old stars and underestimated stellar masses \citep{Shapley2005UVtoMIRobsOfSFGsAtHighz, Pforr2012RecoveringStellarPopPropertiesFromMockFits, Buat2014reliabilityOfSEDfittingAtzgtr1, Michalowski2014massesOfSubmmGalaxiesFavor2cSFHs, Salim2016GSWLC} and also ensures a more realistic range of colors compared to single population models \citep{Pacifici2015importanceOfAppropriateSpectralModelsHighz}.  The inclusion of the second episode of SF is needed to allow for high sSFRs not otherwise attainable by a single exponential with fixed old age \citep[see][]{Ciesla2017mainsequenceWithFlexible2cSFH} and effectively produces rising SFHs, which are known to be more realistic than strictly declining models for many high-redshift galaxies \citep[e.g.,][]{Reddy2012HighzRisingSFHs}. 

We test three different SFH models in addition to the fiducial model.  For the model labelled as `flexible burst' we allow $\tau_{\text{burst}}$ to take on values of 10 Myr, 200 Myr, and 1000 Myr in addition to 20 Gyr.  This allows for a wider range of current SFR at a given mass-weighted age; in other words, for a given burst age, mass fraction, and old e-folding time (which together specify the typical stellar age of the model) different burst e-folding times will lead to different levels of current SF but with similar stellar population ages.  

We include another flexible SFH model for which $\tau_{\text{burst}}$ is again fixed at 20 Gyr but where we allow for burst ages of 10 Myr and 30 Myr, which allows for younger stellar populations compared to the fiducial model; we label this as the `younger flat bursts' model.  Using the flexible burst and younger flat burst models, we will evaluate how the precision of $L_{\text{dust}}$ is related to the stochasticity of the recent SFH.  

Details of the old ($> 1$ Gyr) SFH are poorly constrained by broadband photometry alone.  Nonetheless, we test to see if changing the prescription for the old SFH has noticeable impact on the agreement between $L_{\text{dust}}$ and $L_{\text{TIR}}$.  We use the same grid of parameters as for the fiducial run, but with a delayed exponential SFH instead of a declining exponential; we label this as the `delayed old SFH' model.  The key difference is that for the delayed SFH, the star formation of the old component rises smoothly, peaks, then declines, whereas for exponential the SFH of the old component starts at a maximum value and declines from there.

\subsection{Total IR Luminosities}
\label{Section:BOSA}

To estimate total IR luminosities from the IR photometry, we make use of the IR templates and associated software of \citet{BOSA2021IRTemplates}, referred to as BOSA.  BOSA offers templates based on a sample of 2,584 low-redshift star-forming galaxies (galaxies with AGN have been excluded) and which have extensive coverage in the IR from WISE and Herschel ($12-500 \mu$m).  

The BOSA templates are notable for including a dependence on the specific SFR as well as IR luminosity.  The sSFR dependence helps to ensure the accuracy of $L_{\text{TIR}}$ for high redshift galaxies, for which $L_{\text{TIR}}$ may be biased when using strictly luminosity-dependent templates \citep[see][]{Elbaz2011InfraredMainSequence}.  We note that BOSA is capable of choosing templates based on IR color when multiple IR fluxes are provided, but in the case of a single flux point the templates must also be luminosity-dependent.  The BOSA code requires at least one IR flux point and some estimate of the sSFR as input.  We supply the sSFR from the fiducial SED fitting to obtain $L_{\text{TIR}}$.  We also reproduced all results using a fixed log sSFR of $-9$ (representative of high redshift star-forming galaxies) and find no change in our conclusions.  

We note that 24$\mu$m observed at $z = 1$ corresponds to a rest-frame wavelength of 12$\mu$m while at $z = 2$ the rest-frame wavelength is $8\mu$m.  It has been established that AGN can contribute significantly to galaxy SEDs in the mid-infrared \citep[e.g.,][]{Kirkpatrick2015highredshiftIRtemplates&AGN, Leja2018HotDustSEDfitting...AGNbias}, though unobscured (type 1) AGN can have an effect on the UV-optical-NIR photometry as well \citep[e.g.][]{Wuyts2009SimulatedGalaxiesLessonsForSEDModeling}.  However, in this work we focus mostly on the scatter in $L_{\text{dust}}$ versus $L_{\text{TIR}}$ which will be minimized for the best set of models irrespective of any potential moderate AGN contribution.

\subsection{An assessment of redshift-dependent systematics}
\label{Section:redshiftSystematicOffsetCorrections}

\begin{figure}
    \centering
    \includegraphics[scale=0.6]{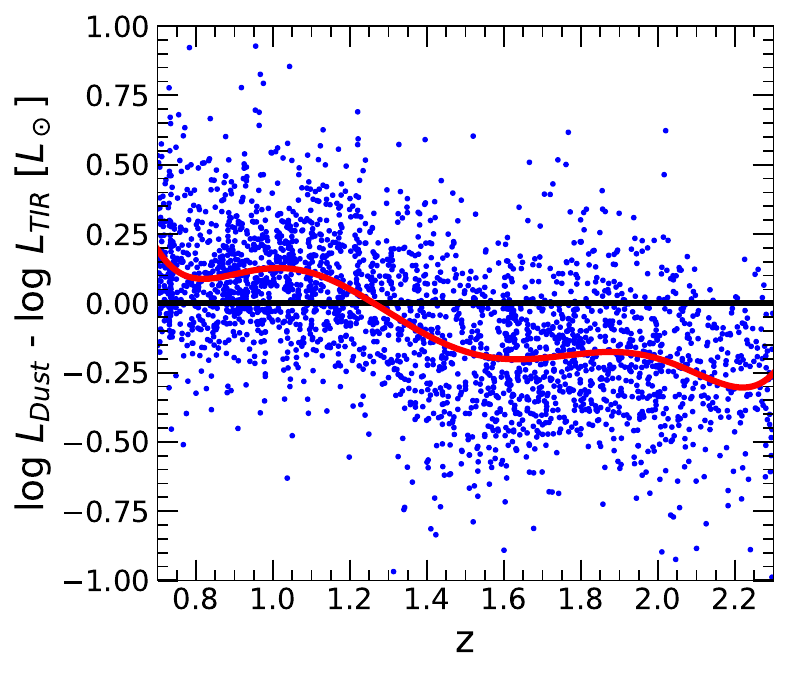}
    \caption{IR luminosity residuals ($L_{\text{dust}} - L_{\text{TIR}}$) versus redshift for the final sample of galaxies.  $L_{\text{dust}}$ is the predicted dust luminosity from the fiducial SED fitting, whereas $L_{\text{dust}}$ is the same quantity determined from 24$\mu$m observations using IR templates. The red line represents a polynomial fit (degree 6) which we use to correct redshift-dependent systematics so we can focus on the strength of the correlation between $L_{\text{dust}}$ and $L_{\text{dust}}$.}
    \label{delta_lir_vs_z}
\end{figure}

\begin{figure}
    \centering
    \includegraphics[scale=0.6]{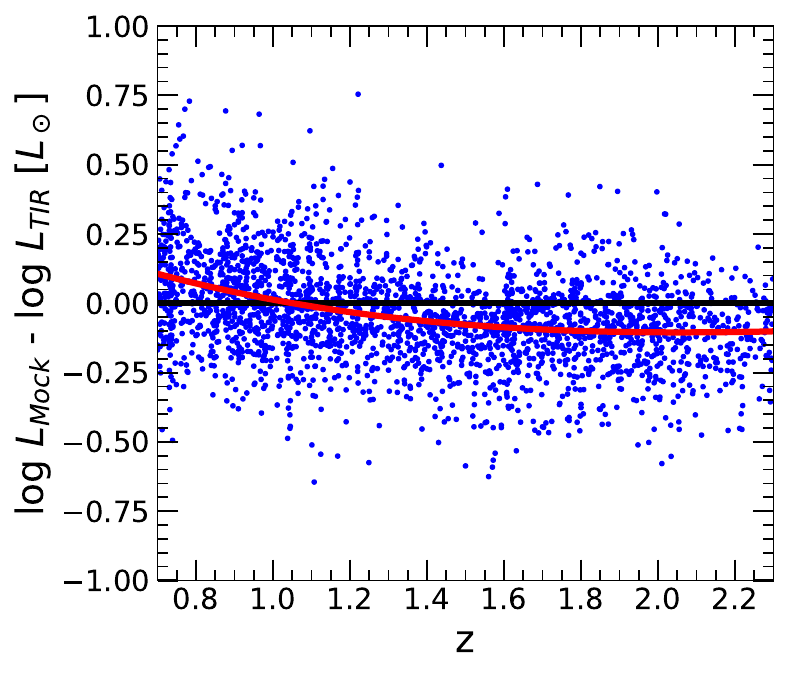}
    \caption{IR luminosity residuals ($L_{\text{dust}} - L_{\text{TIR}}$) versus redshift for the final sample of galaxies. $L_{\text{Mock}}$ represents $L_{\text{dust}}$ from the SED fitting in which real fluxes have been replaced with the with the mock fluxes that must by construction produce $L_{\text{TIR}}$. The red line represents a polynomial fit (degree 6) which helps visualize the redshift-dependent systematics arising from SED fitting alone. These offsets are smaller than the real offsets (Figure \ref{delta_lir_vs_z}) at all redshifts, suggesting that the real offsets may be primarily driven by the systematics in deriving $L_{\text{TIR}}$ from 24$\mu$m observations using IR templates.}
    \label{mock_delta_lirvz}
\end{figure}

Figure \ref{delta_lir_vs_z} shows the difference in $L_{\text{dust}}$ estimated from the fiducial SED fitting run and $L_{\text{TIR}}$ versus redshift. A polynomial fit (degree $= 6$), which is shown in Figure \ref{delta_lir_vs_z} as a red line, helps visualize the typical offset at different redshifts. At $z \lesssim 1.4$, $L_{\text{dust}}$ is typically greater than $L_{\text{TIR}}$ by $\sim 0.2$ dex.  There is a `break' at $z \sim  1.4$ where the systematic offset reverses, beyond which $L_{\text{TIR}}$ is greater than $L_{\text{dust}}$ by $\sim 0.25$ dex.  The `break' at $z \sim 1.4$ corresponds roughly to the redshift where strong PAH features are redshifted into the Spitzer/MIPS $24\mu$m bandpass.  The redshift-dependent offsets between $L_{\text{dust}}$ and $L_{\text{TIR}}$ may be partially due to inflexibility on the part of the sSFR-dependent templates to simultaneously account for the variation in PAH features as well. Unidentified systematics in the SED fitting may also contribute to the offset \citep[e.g.,][]{Leja2019OlderMoreQuiescentUniverse}.  

To try and discern whether the redshift-dependent offsets of Figure \ref{delta_lir_vs_z} are due primarily to the SED fitting or to the IR templates, we perform a mock SED fitting test, which aims to determine how well can SED fitting (still using only UV-near-IR constraints) recover `known' IR luminosities . To do so, we first repeat the fiducial SED fitting (see Section \ref{Section:FiducialModelPriors}), but include $L_{\text{TIR}}$ as a strong constraint. Choosing very small errors ($1\%$ of $L_{\text{TIR}}$) guarantees that the resulting $L_{\text{dust}}$ is equal to $L_{\text{TIR}}$. The best-fit fluxes from this IR-constrained fitting now have to conform with this $L_{\text{TIR}}$.  For the mock fitting, the mock fluxes are drawn from a Gaussian centered on the best-fit flux assuming the same signal-to-noise as the original observations (so mock error $=$ (observed error / observed flux) $\times$ best-fit flux). The mock fitting is performed on the mock fluxes and mock errors, using fiducial models and with no IR constraints. Any differences in the mock $L_{\text{dust}}$, which we refer to as $L_{\text{Mock}}$, and $L_{\text{TIR}}$ will then be entirely attributable to the SED fitting. 

Figure \ref{mock_delta_lirvz} shows $L_{\text{Mock}} - L_{\text{TIR}}$ versus redshift. As with Figure \ref{delta_lir_vs_z}, we show a polynomial fit (degree $= 6$) as a red line. Comparing to Figure \ref{delta_lir_vs_z}, we see that the differences obtained from the mocks are smaller than the real differences at all redshifts, and lack most of the inflections present in the real data. Thus we conclude that the redshift-dependent systematic offsets between the actual $L_{\text{dust}}$ and $L_{\text{TIR}}$ are partially due to limitations in the IR templates and partially arising from the SED fitting. The latter are on the order of 0.1 dex, which should not preclude us from addressing the goals of this study. 

The goal of this work is to identify best practices for UV-optical-NIR SED fitting using the scatter in $L_{\text{dust}}$ versus $L_{\text{TIR}}$, rather than the bulk offsets. We therefore correct for the redshift-dependent systematics using a polynomial fit (degree $= 6$), which is shown in Figure \ref{delta_lir_vs_z} as a red line, to place $L_{\text{dust}}$ and $L_{\text{TIR}}$ on a more equal footing.  For each galaxy, we subtract from $L_{\text{dust}}$ the value of the polynomial at that galaxy's redshift.  The same correction (based on fiducial $L_{\text{dust}}$) is applied to all estimates of $L_{\text{dust}}$.

\section{Results}
\label{Section:Results}

In this section, we test variations on a fiducial SED model in order to evaluate the level of agreement between the dust luminosity ($L_{\text{dust}}$) predicted from the UV-optical-NIR fitting and the $L_{\text{TIR}}$ inferred from the Spitzer 24$\mu$m photometry.  We aim to determine which treatments for dust, metallicity, choice of SPS models, and SFH maximize the degree of correlation between $L_{\text{TIR}}$ and $L_{\text{dust}}$.  We also explore systematics in the estimated stellar mass (log $M_*$) across different models.  

While the goal of the study is to arrive at more precise SFRs, comparing $L_{\text{dust}}$ and $L_{\text{TIR}}$ instead of SFRs is advantageous for two reasons.  First, $L_{\text{TIR}}$ includes emission from dust heated from stellar populations of all ages, so we do not need to isolate only the young populations when deriving an SFR estimate from IR emission.  Second, on the SED fitting side, the SFR needs to be averaged on some timescale, and it is not obvious which timescale is most appropriate when comparing to SFRs obtained from $L_{\text{TIR}}$ (or $L_{\text{TIR}}$ in combination with $L_{\text{UV}}$ in the popular hybrid method).  On the other hand, if $L_{\text{dust}}$ determined from the stellar continuum SED fitting agrees with $L_{\text{TIR}}$, then the SED fitting SFRs should be robust too.

To evaluate agreement we primarily look at the standard deviation ($\sigma$) of the difference between $L_{\text{dust}}$ and $L_{\text{TIR}}$ and not the offsets (which are set to zero for the fiducial run by definition, see Section \ref{Section:redshiftSystematicOffsetCorrections}). For stellar mass comparisons we discuss primarily the mean relative offset ($\Delta$) between various SED fitting schemes and the fiducial one, since we have no stellar masses that can serve as a ground-truth.

\subsection{Dust attenuation}

\begin{figure}
    \centering
    \gridline{
          \fig{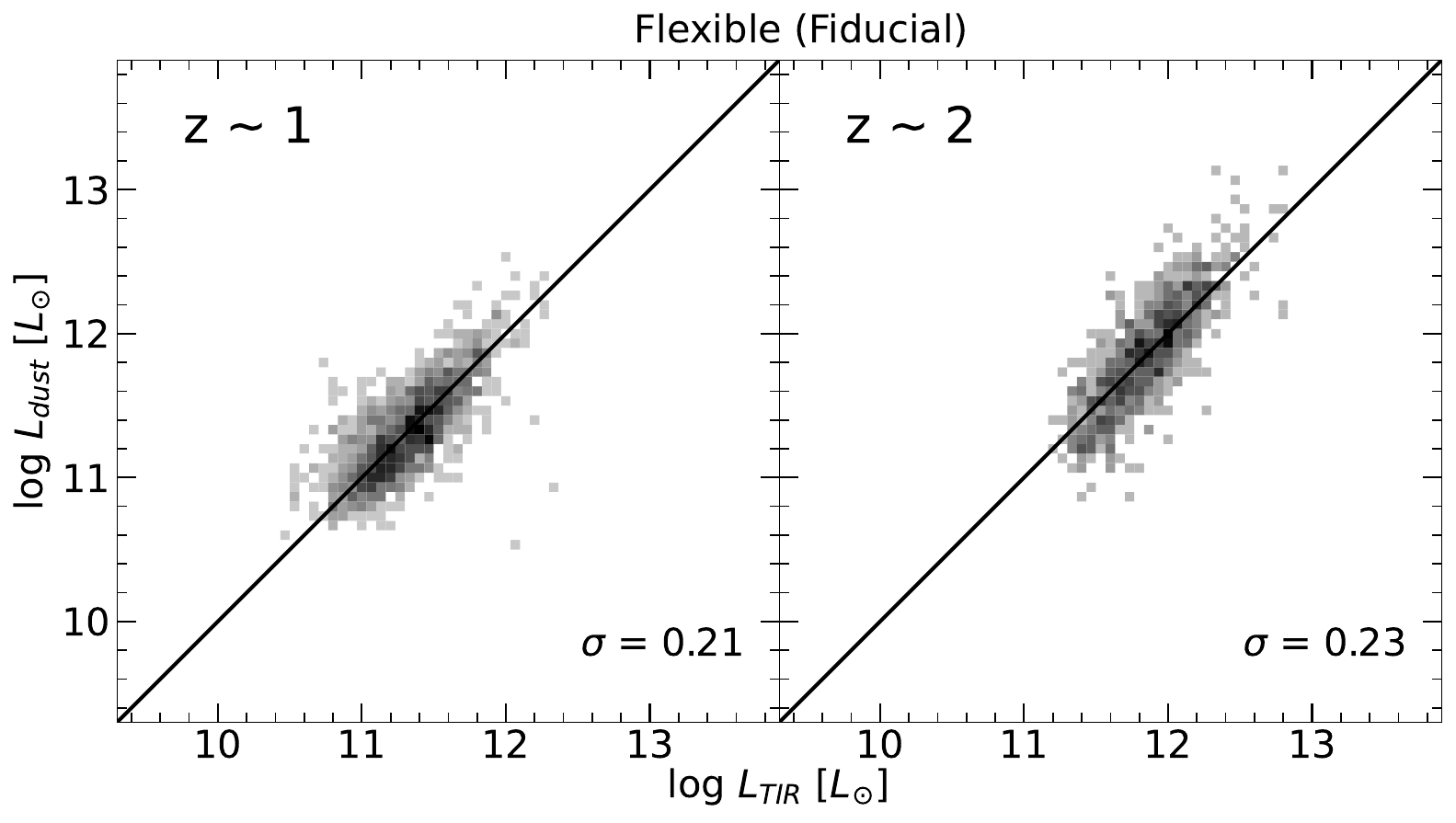}{0.45\textwidth}{}
    }
    \vspace*{-\baselineskip}
    \gridline{
          \fig{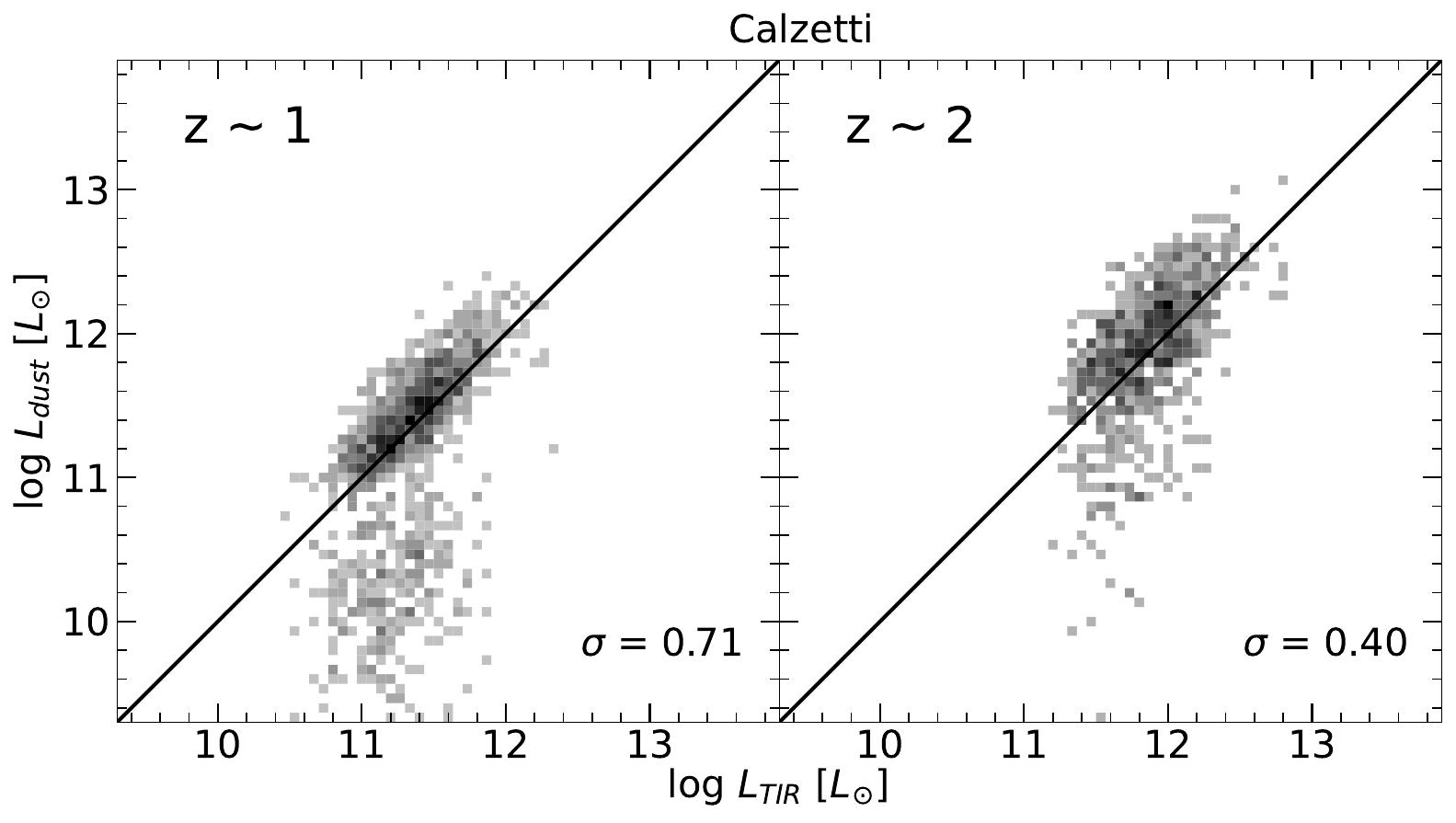}{0.45\textwidth}{}
    }
    \vspace*{-\baselineskip}
    \gridline{
          \fig{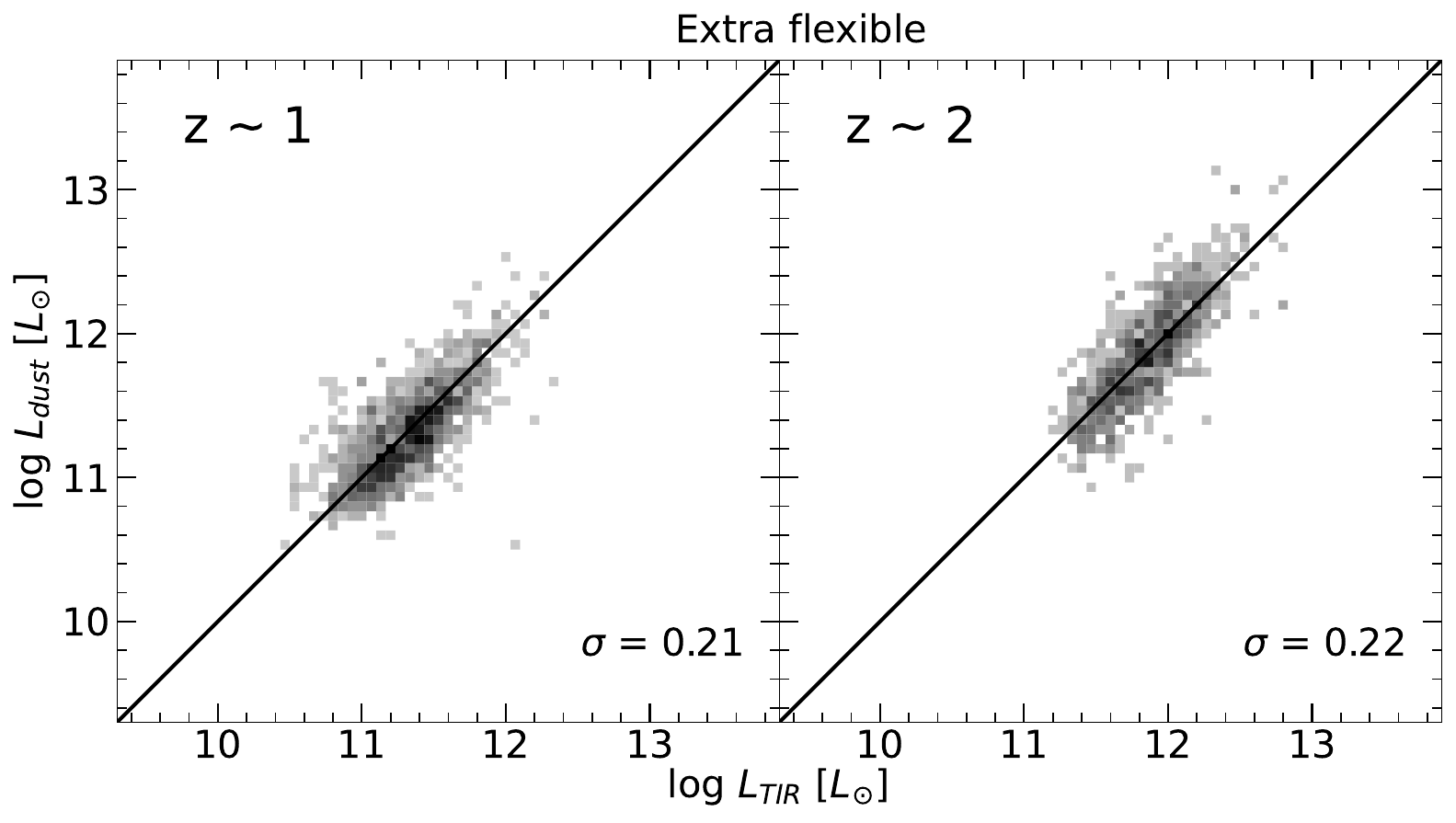}{0.45\textwidth}{}
    }
    \vspace*{-\baselineskip}
    \gridline{
    \fig{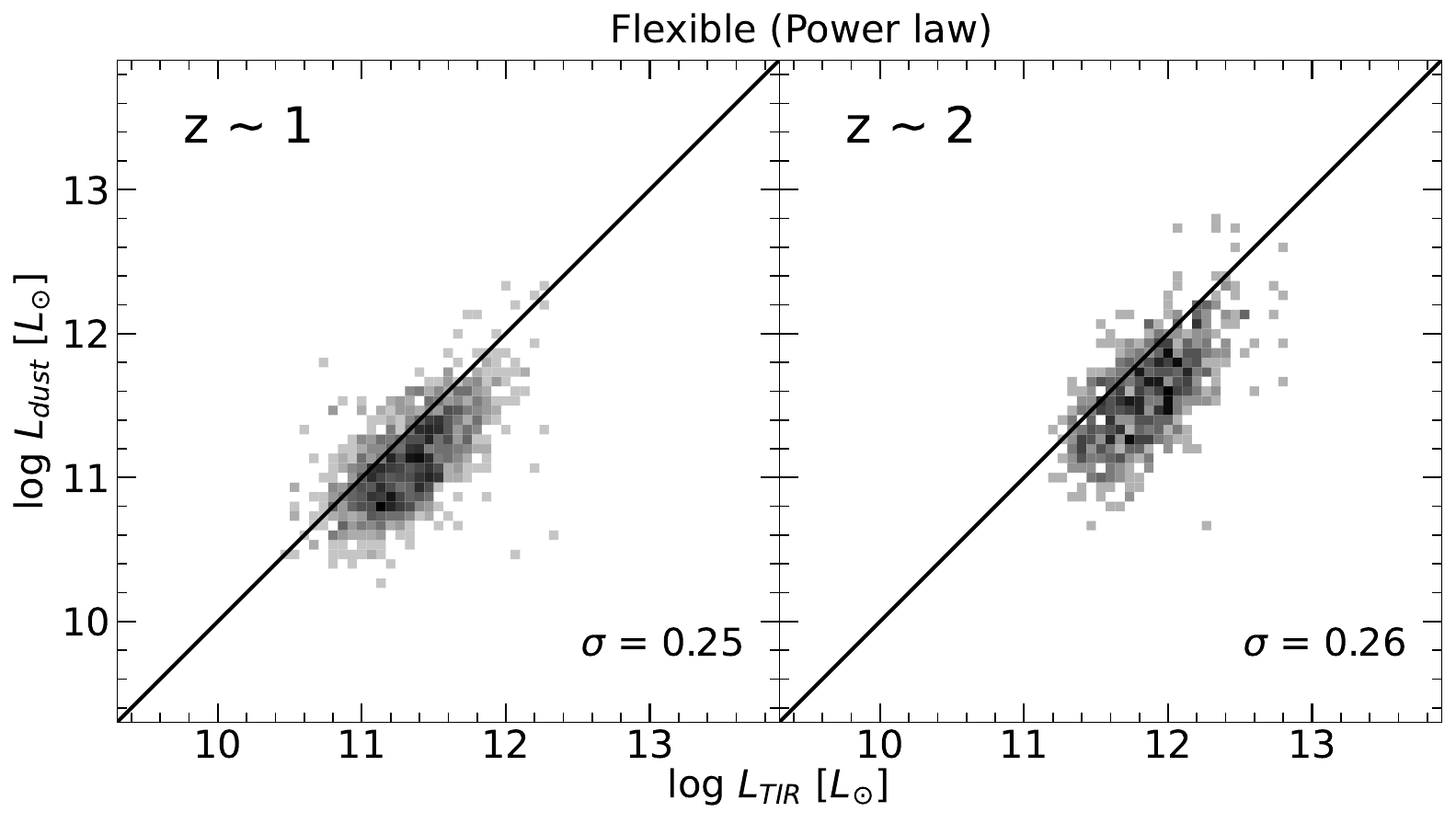}{0.45\textwidth}{}
    }
    \vspace*{-\baselineskip}
    \caption{Comparison of $L_{\text{dust}}$ from UV-optical-NIR SED fitting to $L_{\text{TIR}}$ for different dust attenuation models. The fiducial model outperforms both simpler and more complex models, as well as the alternative power law model.  In this and similar subsequent panels, the scatter $\sigma$ is the standard deviation in log $L_{\text{dust}} -$ log $L_{\text{TIR}}$.}
    \label{DustRunPlots}
\end{figure}

We first explore how variations in the assumed dust attenuation law affect the correlation between $L_{\text{dust}}$ and $L_{\text{TIR}}$.  Figure \ref{DustRunPlots} shows the $L_{\text{TIR}}$ and $L_{\text{dust}}$ comparison for the fiducial model (top panel) alongside the comparisons for the Calzetti (second panel), extra flexible (third panel), and flexible power law (bottom panel) models. 

The Calzetti model has a single slope for the attenuation curve, allows no UV bump, and assumes the same attenuation for young and old stars, unlike the fiducial model which varies the slope and bump strength and applies extra attenuation to young stars. The Calzetti model results in strongly underestimated $L_{\text{dust}}$ among galaxies with relatively low $L_{\text{TIR}}$, which is largely attributable to the absence of steeper slopes.  Omitting this low $L_{\text{dust}}$ tail, the rest of the galaxies agree with $L_{\text{TIR}}$ relatively well, though the scatter is still higher at $z\sim2$ than for the fiducial dust model. 

The extra flexible model allows the attenuation curves applied to old and young stars to vary independently from one another, unlike for fiducial where the young and old slopes are kept the same.  For extra flexible attenuation, there is a slight decrease in scatter at $z \sim 2$.  Overall, the extra flexible model shows only a marginal improvement over the flexible fiducial model while being much more computationally expensive due to the expanded grid, thus we do not recommend its use.  Assuming a flexible power law parameterization instead of the fiducial model significantly increases the scatter with respect to the fiducial run at both redshift windows.

\subsection{Stellar population synthesis models}

\begin{figure}
    \centering
    \gridline{\fig{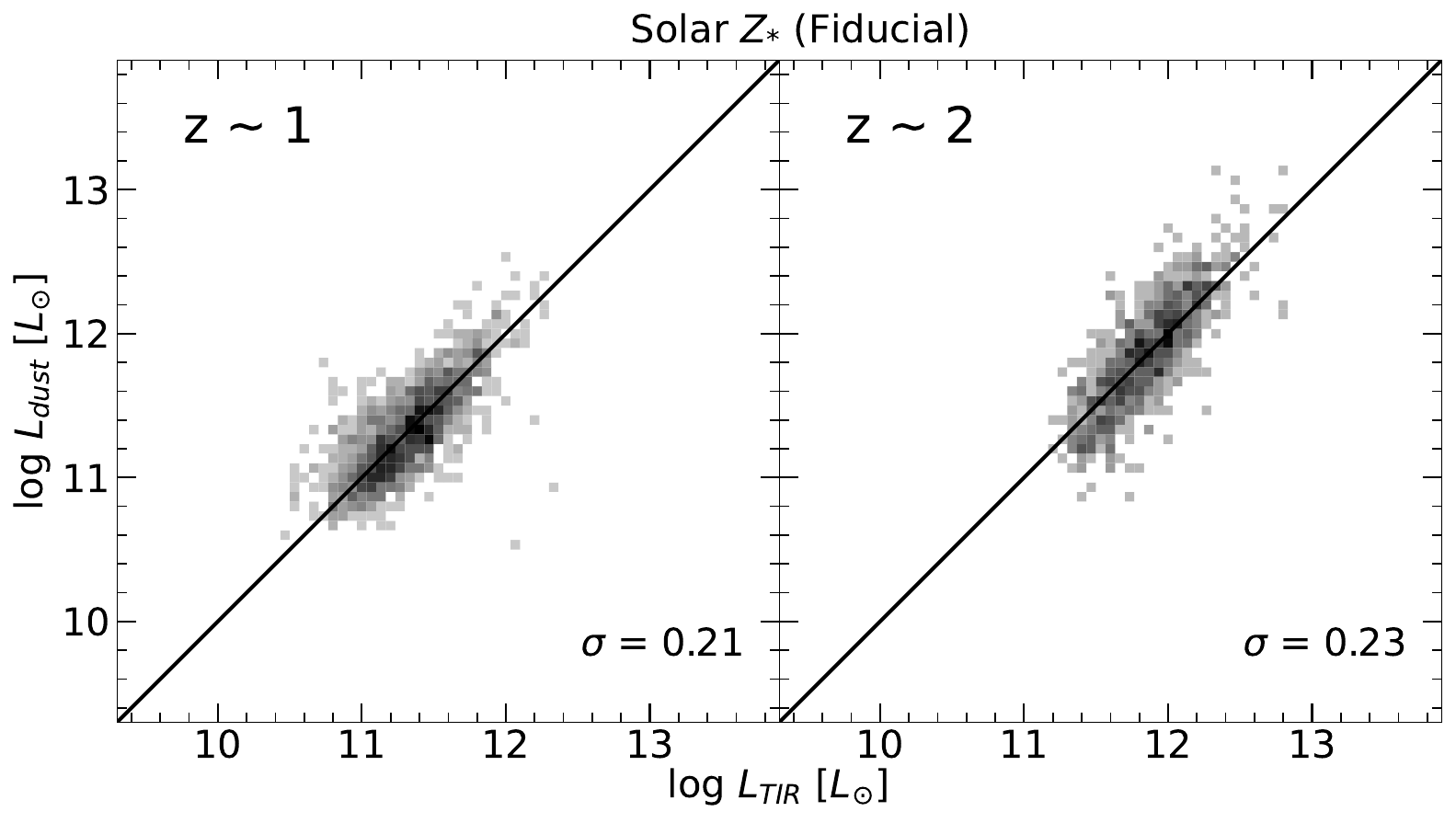}{0.45\textwidth}{}
          }
    \vspace*{-\baselineskip}
    \gridline{\fig{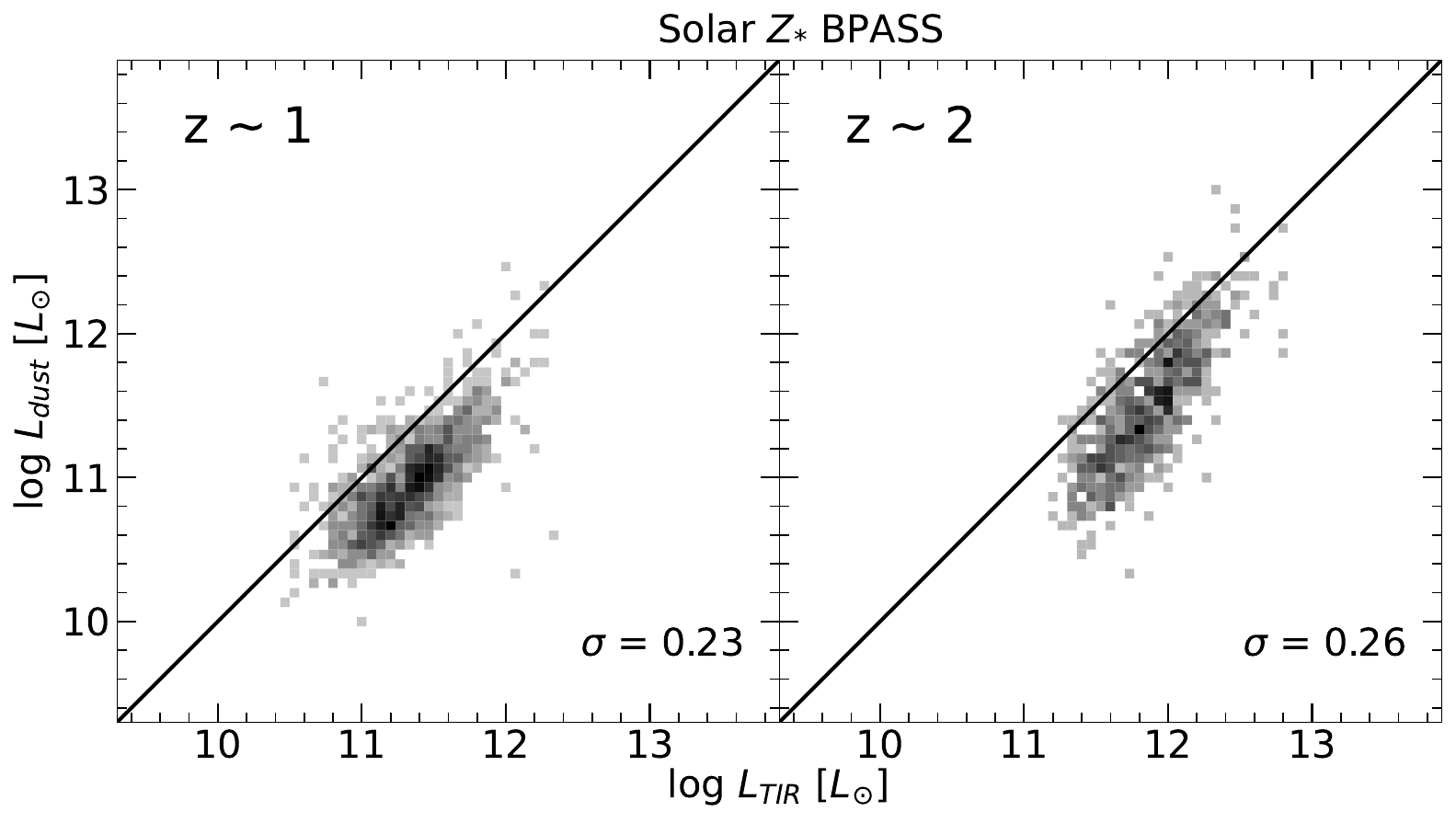}{0.45\textwidth}{}
          }
    \vspace*{-\baselineskip}
    \caption{Comparison of $L_{\text{dust}}$ from UV-optical-NIR SED fitting $L_{\text{TIR}}$ for different stellar population models.  BC03 models are favored compared to BPASS models.  BPASS models predict systematically lower $L_{\text{dust}}$ compared to BC03 at both redshifts.}
    \label{SPSRunPlots}
\end{figure}

We now consider how varying assumptions about the stellar populations impact the precision of $L_{\text{dust}}$.  The top panel of Figure \ref{SPSRunPlots} shows results for the fiducial solar $Z_*$ BC03 model, while the bottom panel shows results for the solar $Z_*$ BPASS model.  We see that using BPASS SPS models instead of the fiducial BC03 SPS models results in increased scatter between $L_{\text{dust}}$ and $L_{\text{TIR}}$ at both redshifts.  BPASS models appear to predict systematically lower $L_{\text{dust}}$ compared to BC03 at both redshifts, despite the stronger ionizing flux of BPASS models compared to BC03 models for a given stellar age.  The lower values of $L_{\text{dust}}$ may be due to the lower stellar masses predicted by BPASS (see Figure \ref{MassPlots}), resulting in fewer stars and less overall heating.  There is also a much more prominent non-linearity in $L_{\text{dust}}$ versus $L_{\text{TIR}}$ for BPASS models at $z \sim 2$.

\subsection{Stellar metallicity}

\begin{figure}
    \centering
    \gridline{\fig{BOSA24_vs_SED_SPSlabel.pdf}{0.45\textwidth}{}
          }
    \vspace*{-\baselineskip}
    \gridline{\fig{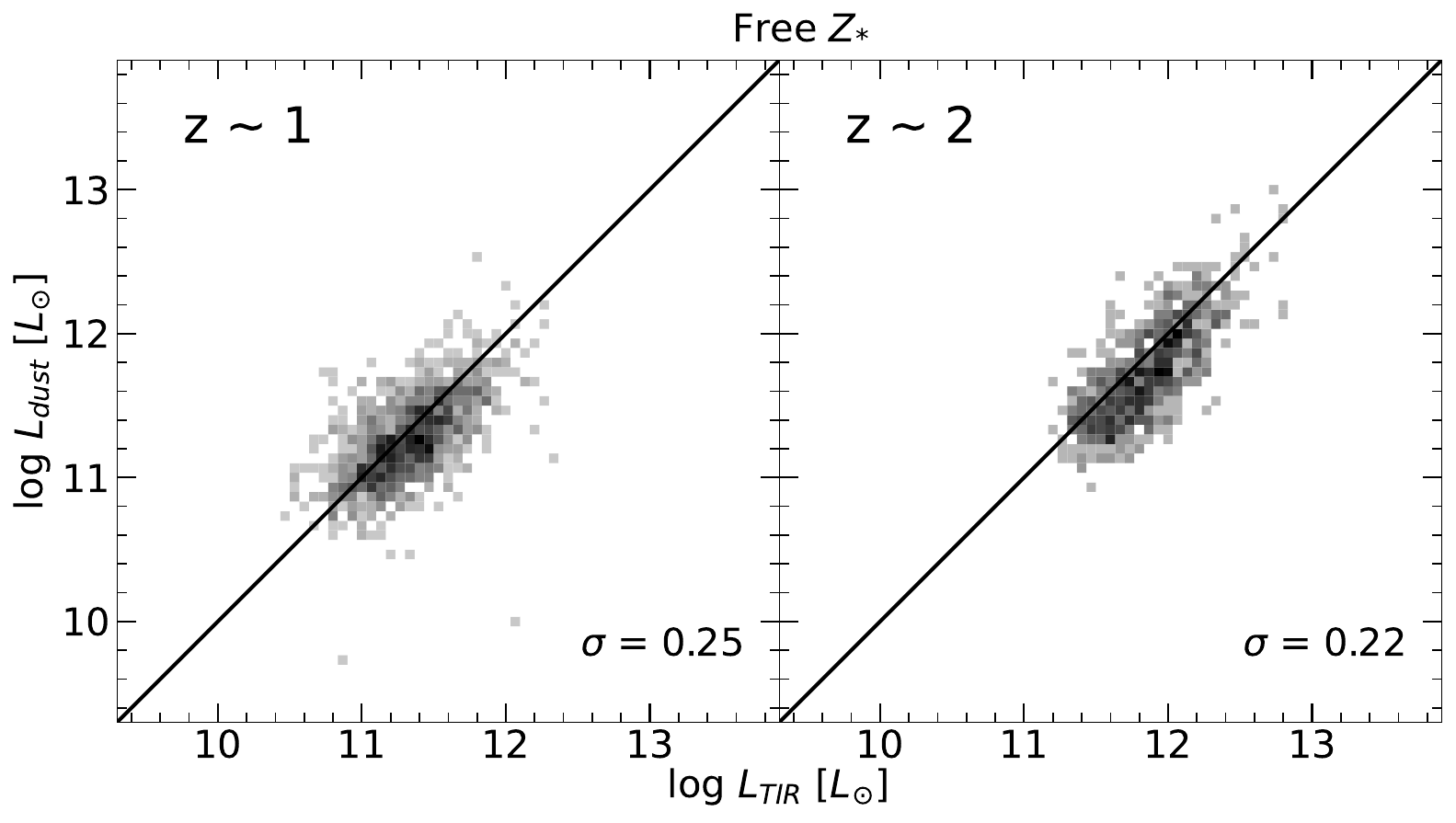}{0.45\textwidth}{}
          }
    \vspace*{-\baselineskip}
    \gridline{\fig{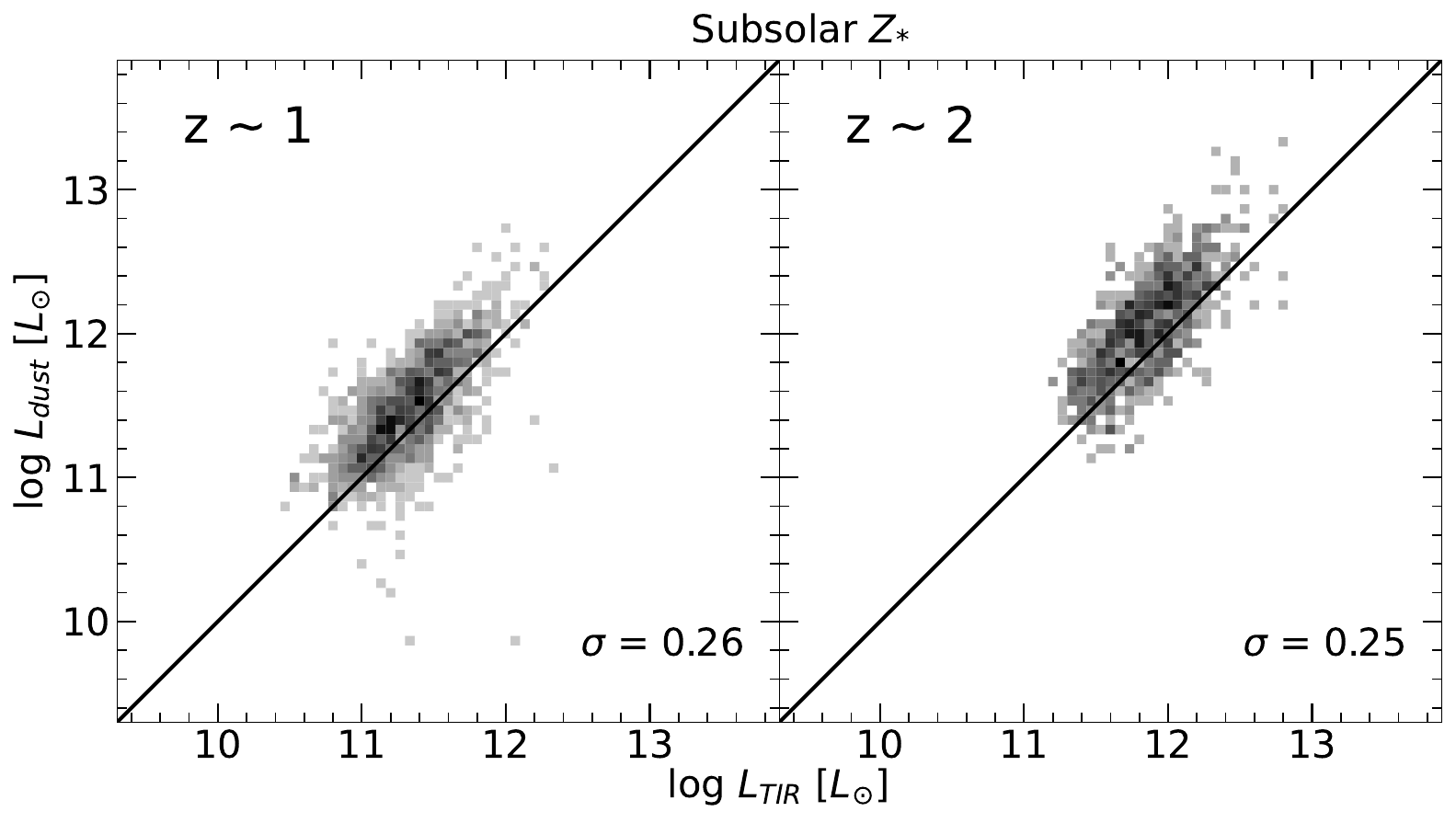}{0.45\textwidth}{}
          }
    \vspace*{-\baselineskip}
    \caption{Comparison of $L_{\text{dust}}$ from UV-optical-NIR SED fitting $L_{\text{TIR}}$ for different stellar metallicity models. For fixed metallicity, the choice of metallicity is a major systematic. The fiducial model, which fixes the metallicity at solar, is favored overall. Free metallicity models may be slightly disfavored at $z \sim 1$ due to dust-metallicity degeneracy, while at $z \sim 2$ the signal-to-noise of observations has a greater impact on the uncertainty of $L_{\text{dust}}$ than the degeneracy.}
    \label{MetalRunPlots}
\end{figure}

We now consider how varying assumptions about the stellar metallicity impacts the precision of $L_{\text{dust}}$.  We show the comparisons for the different models in Figure \ref{MetalRunPlots}.  The fiducial model, which fixes the metallicity at solar ($Z_* = 0.02$), is shown in the top panel.  Allowing the stellar metallicity to vary (Free $Z_*$, middle panel) results in increased scatter compared to the fiducial model at $z\sim 1$, which is likely attributable to dust-metallicity degeneracy, but marginally decreased scatter at $z \sim 2$. The slightly lower scatter at $z \sim 2$ for free metallicity models may suggest that the uncertainties in $L_{\text{dust}}$ at higher redshifts are dominated by the photometric signal-to-noise rather than modeling uncertainties. The net scatter is lower for the fiducial model compared to the free $Z_*$ model, so we find the fiducial model to be favored overall. 

While a fixed solar metallicity is often assumed in SED fitting for simplicity, it is not obviously the best choice for galaxies at high redshift, in particular at $z \sim 2$ where the gas-phase metallicity is lower than at $z \sim 0$ by as much as a factor of two \citep[see e.g.,][]{Maiolino2008MassMetallicityRelationHighz, Wuyts2016evolutionOfMetallicityAtIntermediateRedshifts}.  However, we find that for the subsolar $Z_*$ model (bottom panel), which fixes the metallicity at $Z_* = 0.008$ (or 40\% solar), the scatter is increased compared to the fiducial model and there is a systematic shift to higher $L_{\text{dust}}$; the systematic shift seems to occur because, with less metallicity, the dust attenuation must increase to match the observed colors.  Interestingly, the opposite occurs when the metallicity is fixed at supersolar ($Z_* = 0.05$), which causes a systematic shift to \textit{lower} $L_{\text{dust}}$.  We also do not find any improvement over the fiducial model at either redshift when the metallicity is fixed at even lower values (e.g., $Z_{*} = 0.004$).

\subsection{Star Formation History}

\begin{figure}
    \centering
    \gridline{
          \fig{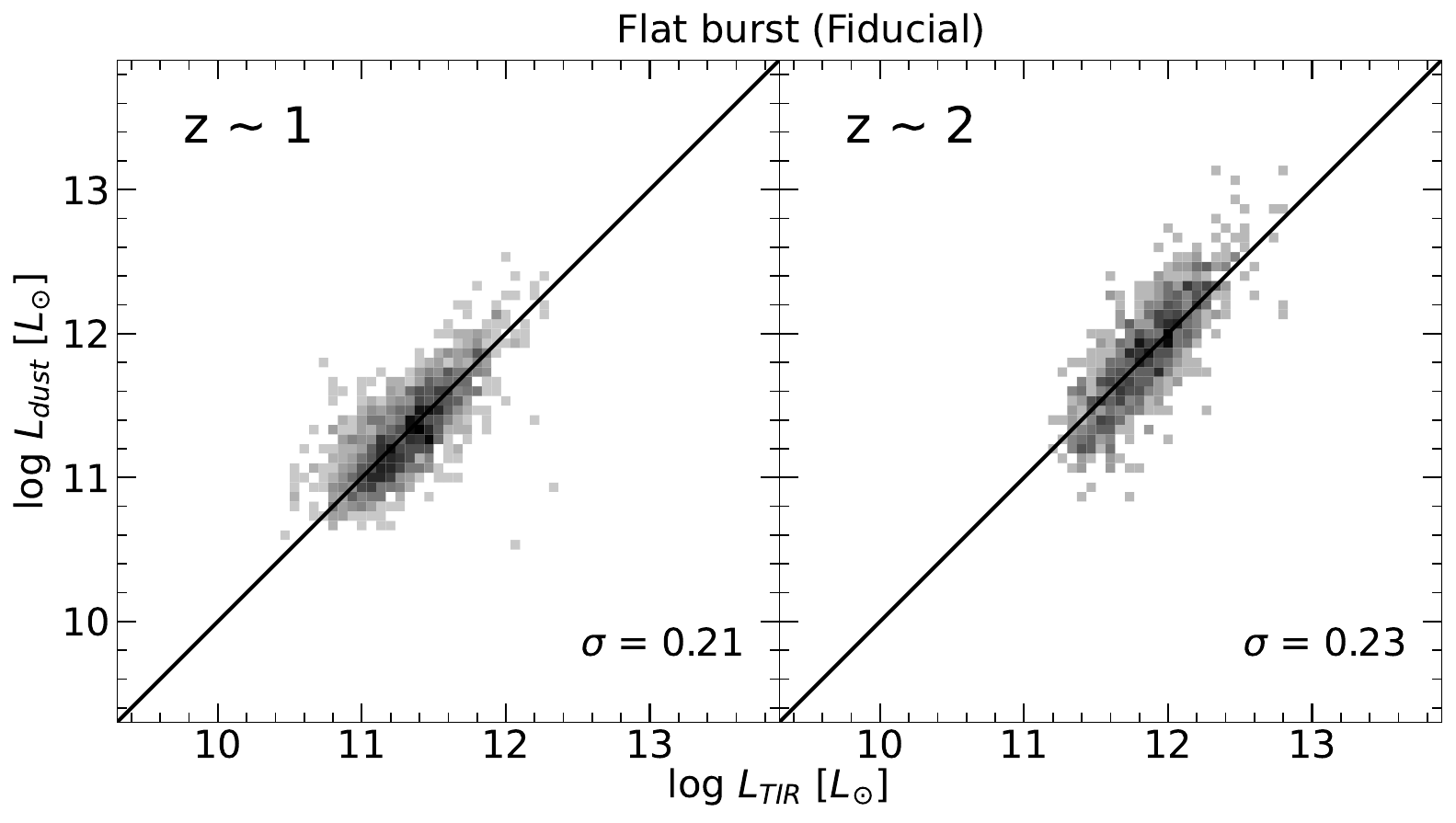}{0.45\textwidth}{}
    }
    \vspace*{-\baselineskip}
    \gridline{
          \fig{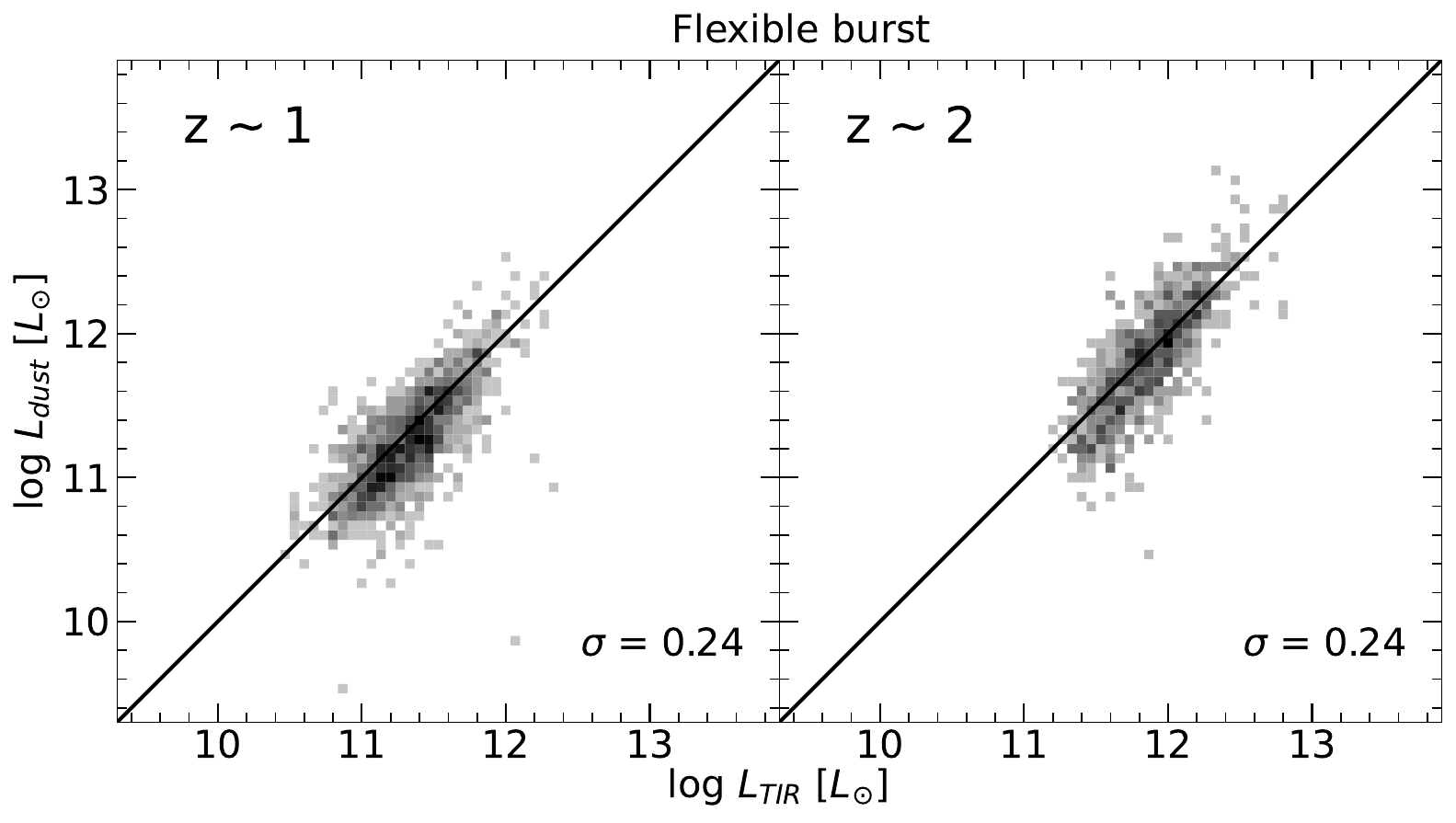}{0.45\textwidth}{}
    }
    \vspace*{-\baselineskip}
    \gridline{
          \fig{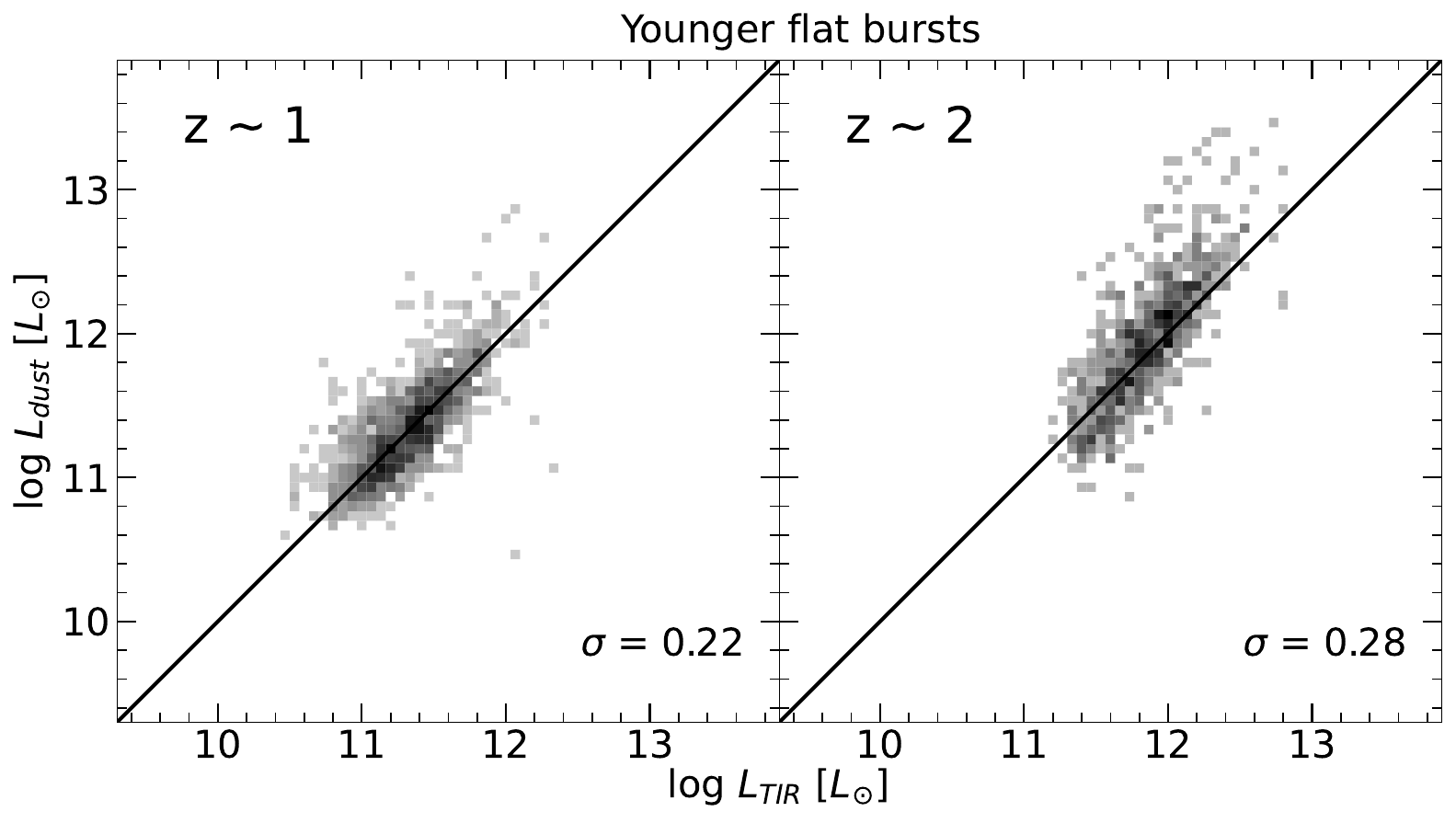}{0.45\textwidth}{}
    }
    \vspace*{-\baselineskip}
    \gridline{
          \fig{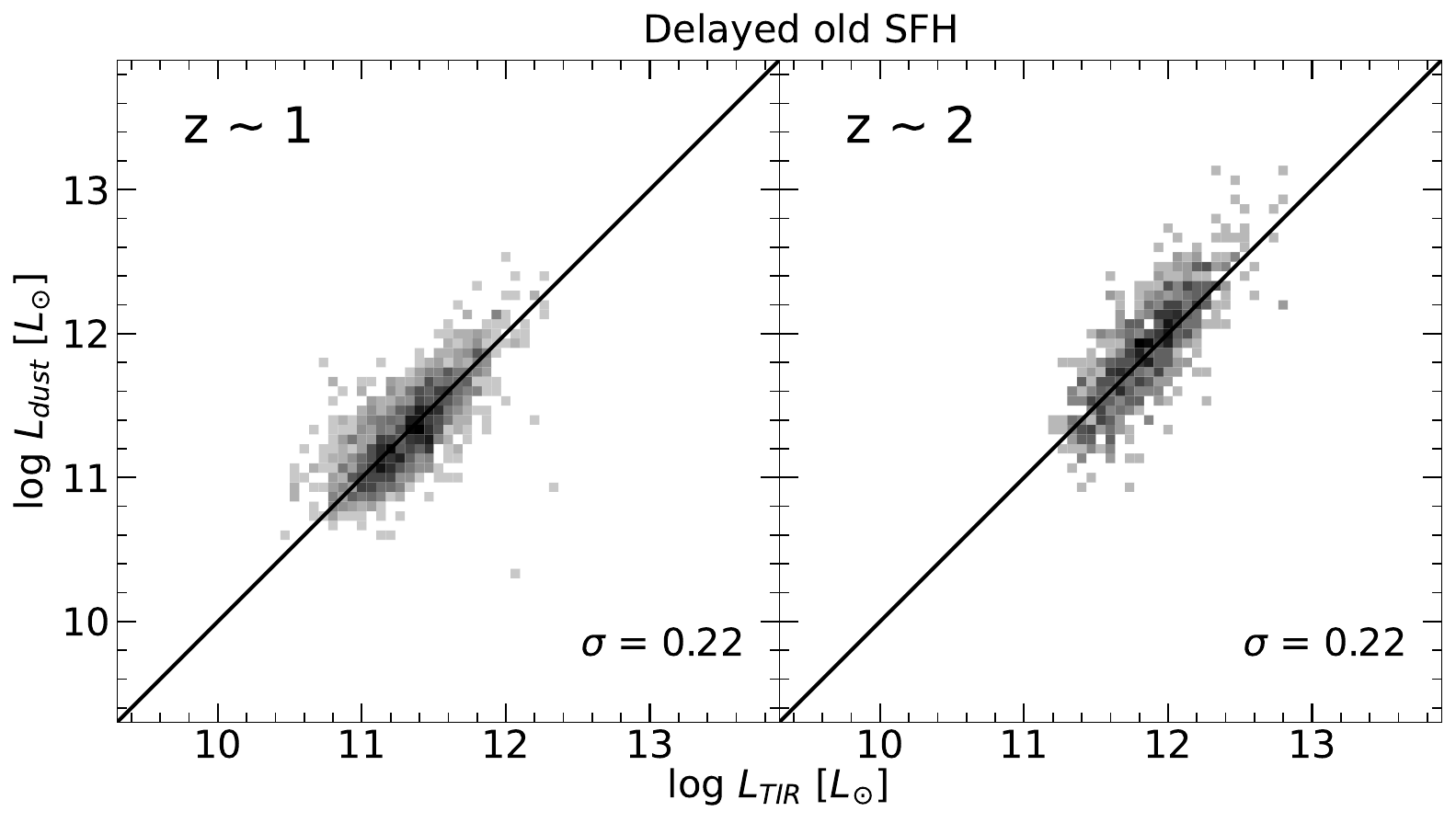}{0.45\textwidth}{}
    }
    \vspace*{-\baselineskip}
    \caption{Comparison of $L_{\text{dust}}$ from UV-optical-NIR SED fitting to $L_{\text{TIR}}$ for different types of star formation history models.  The fiducial model outperforms other models that allow for more variability in the recent SFH. Adopting a delayed exponential parameterization for the old component, instead of an exponential form, has little effect.}
    \label{SFHRunPlots}
\end{figure}

Finally, we explore how changing the type of SFH model affects the precision of $L_{\text{dust}}$.  Figure \ref{SFHRunPlots} shows the fiducial model (top panel), the flexible burst model (second panel), the younger flat bursts model (third panel), and the delayed old SFH model (bottom panel).  The fiducial model is a two-component parameterization featuring a declining exponential old component with fixed age superimposed with a second episode of recent SF with essentially constant SFR and varying mass fraction.  We find that the flexible burst model, which differs from fiducial in that the burst is allowed to decline  (variable $\tau_{\text{burst}}$) instead of remaining flat, increases the scatter compared to the fiducial model at both redshifts.  The increase in scatter for the flexible burst model is most significant for relatively low $L_{\text{TIR}}$ galaxies at $z \sim 1$.

The younger flat bursts model, which differs from fiducial in that it allows for burst ages younger than 100 Myr (10 Myr and 30 Myr), significantly increases the scatter among the highest $L_{\text{TIR}}$ galaxies (which tend to be fit with these younger bursts) compared to fiducial. Finally, we consider the model that treats the old component as the delayed exponential.  We find only marginally increased scatter at $z \sim 1$ and marginally decreased scatter at $z \sim 2$ compared to the fiducial model; therefore, there is little practical difference between the two, i.e., both represent good choices. 

\subsection{Stellar masses}

\begin{figure*}
    \centering
    \gridline{
          \fig{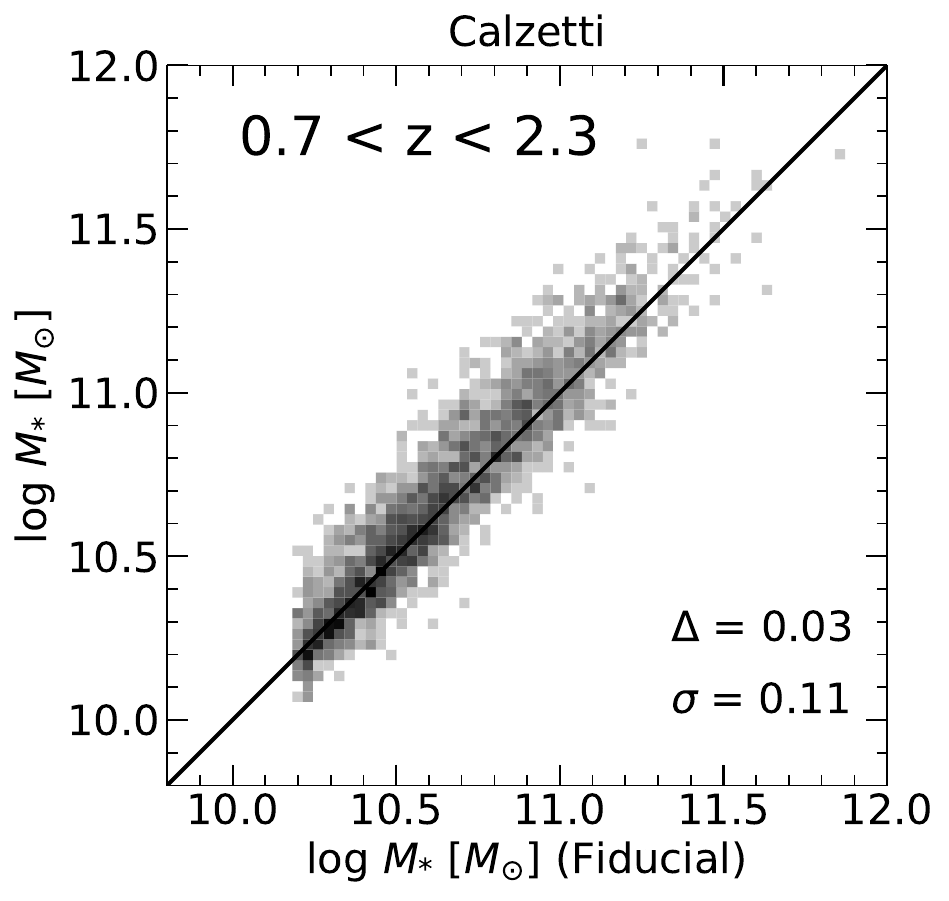}{0.3\textwidth}{}
          \fig{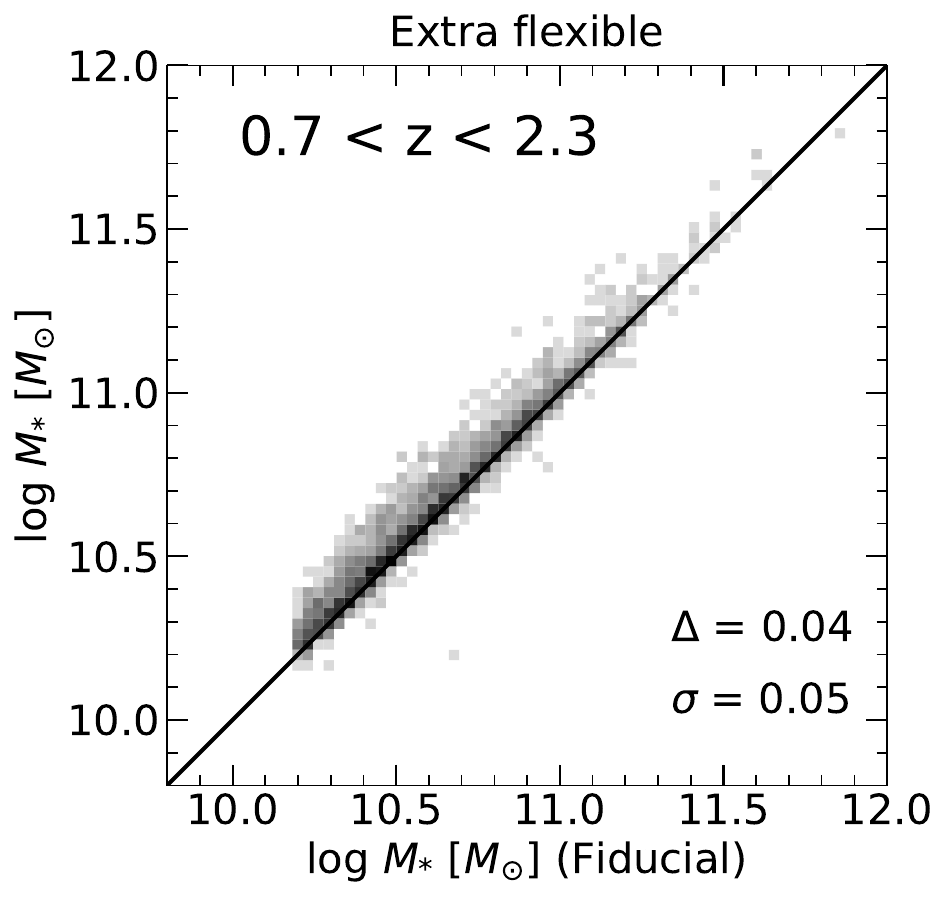}{0.3\textwidth}{}
          \fig{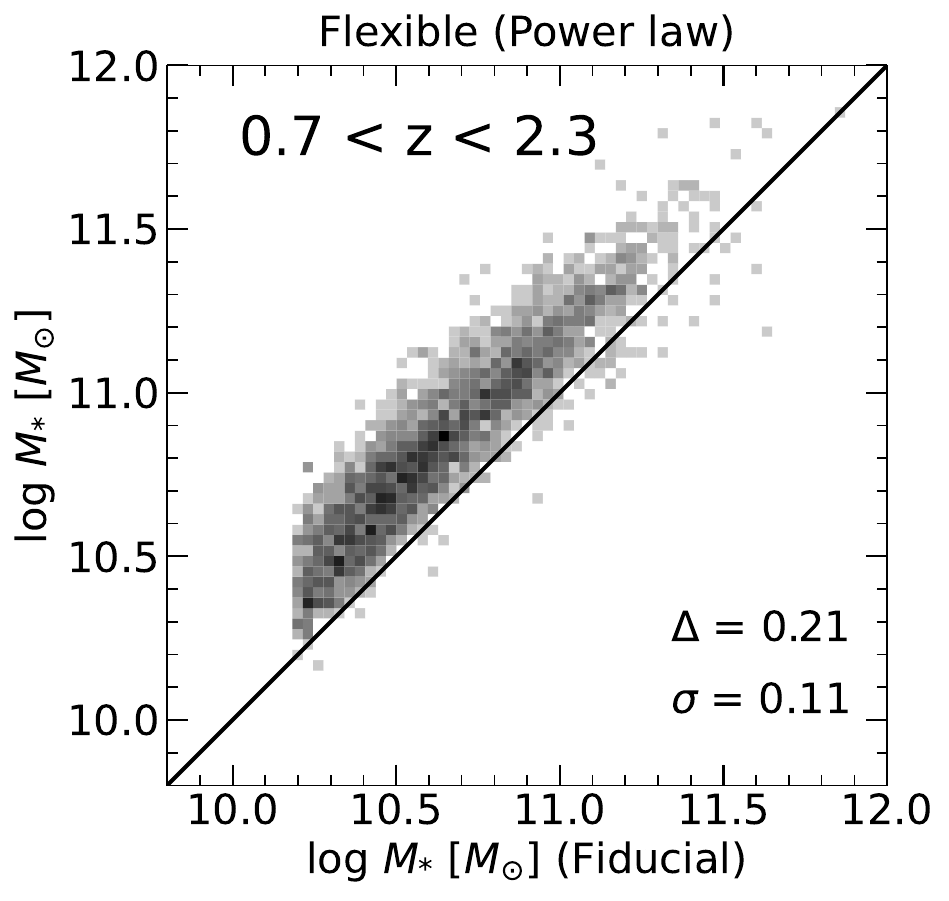}{0.3\textwidth}{}
          }
    \gridline{
          \fig{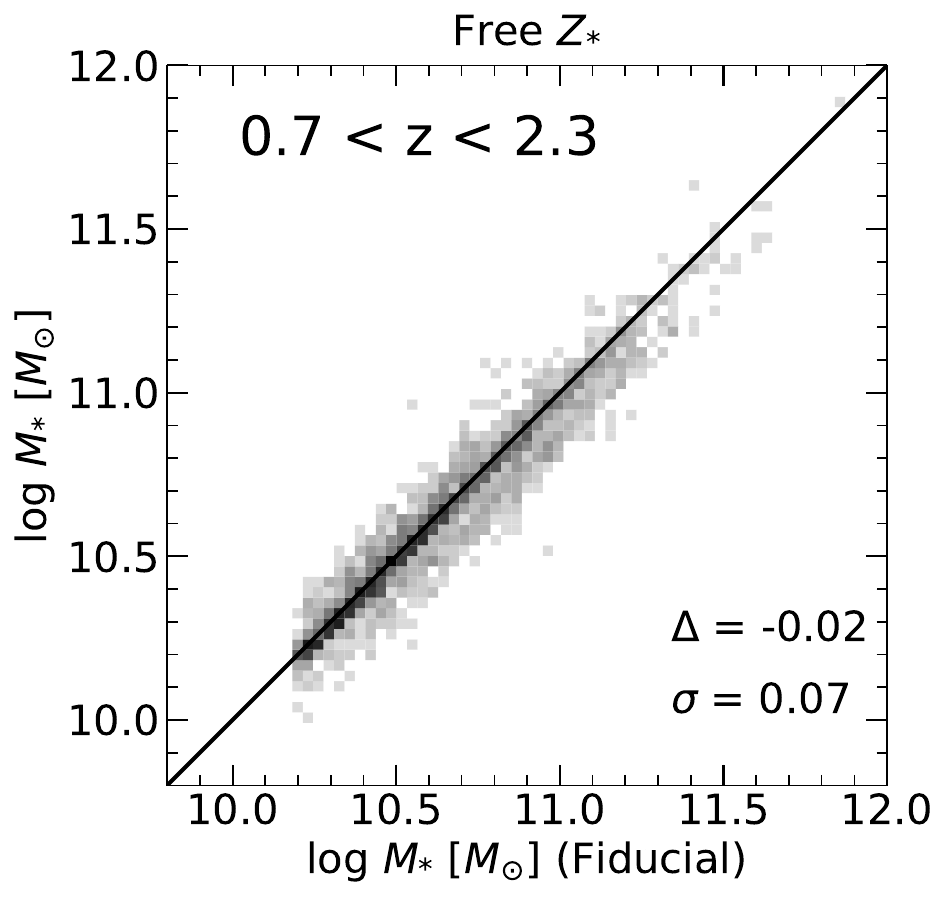}{0.3\textwidth}{}
          \fig{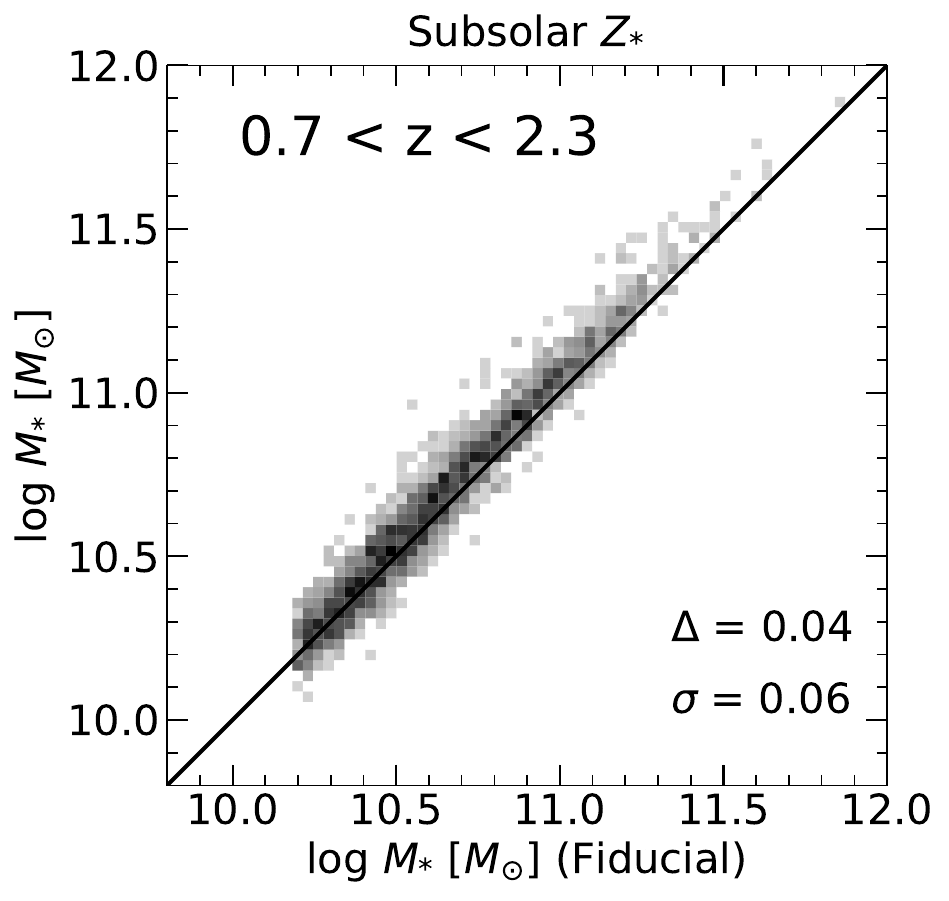}{0.3\textwidth}{}
          \fig{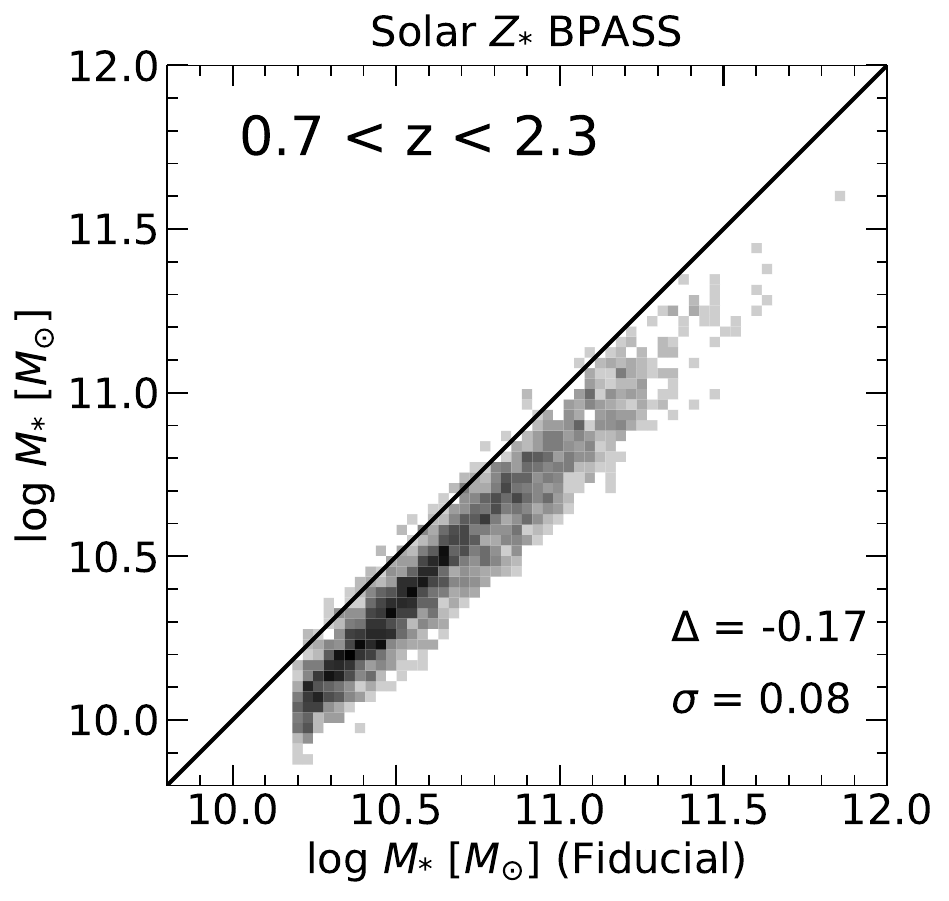}{0.3\textwidth}{}
          }
    \gridline{
          \fig{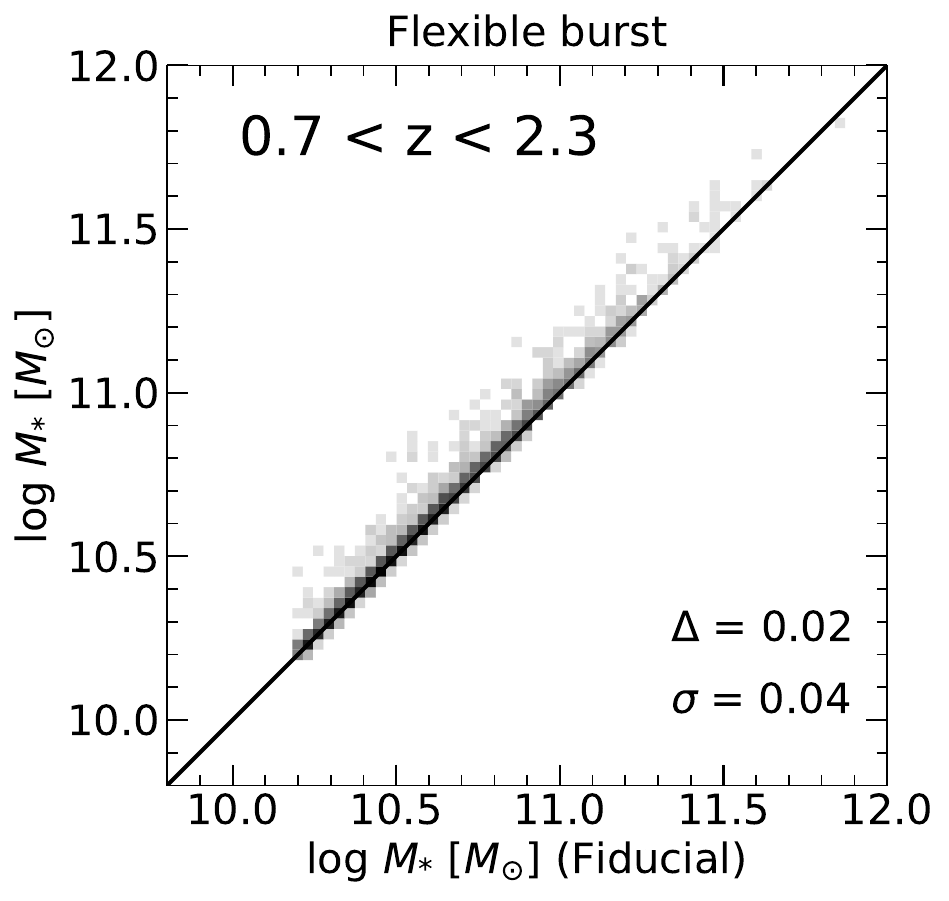}{0.3\textwidth}{}
          \fig{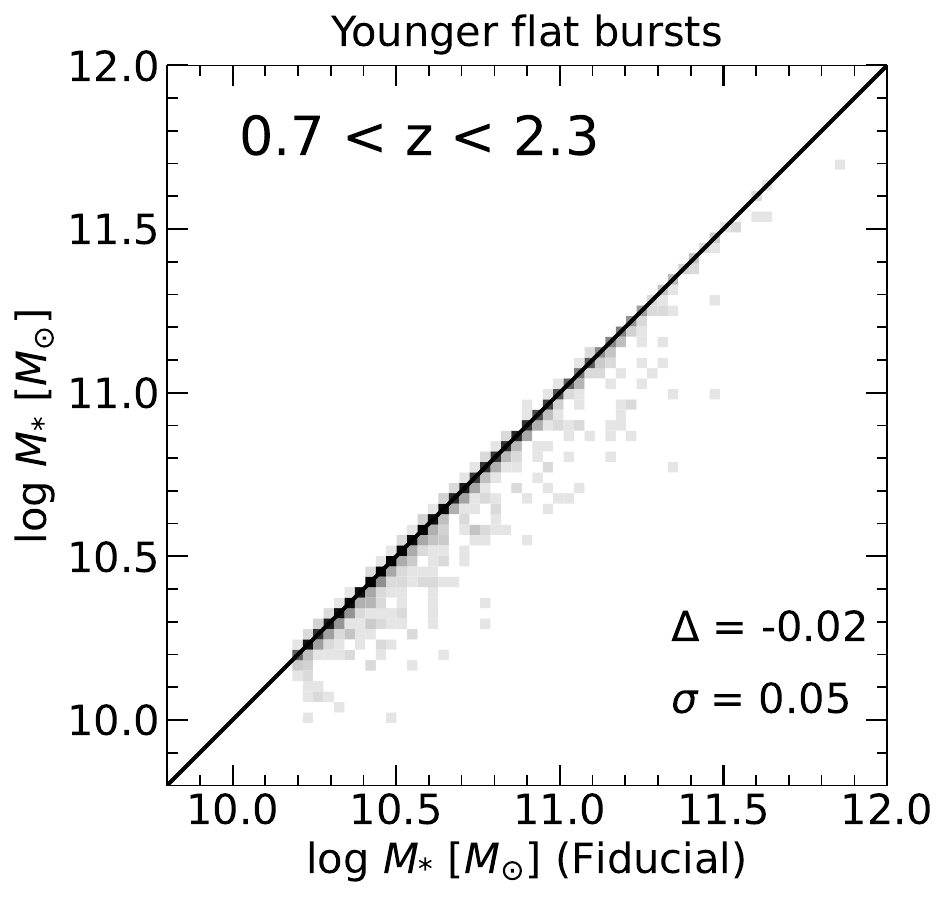}{0.3\textwidth}{}
          \fig{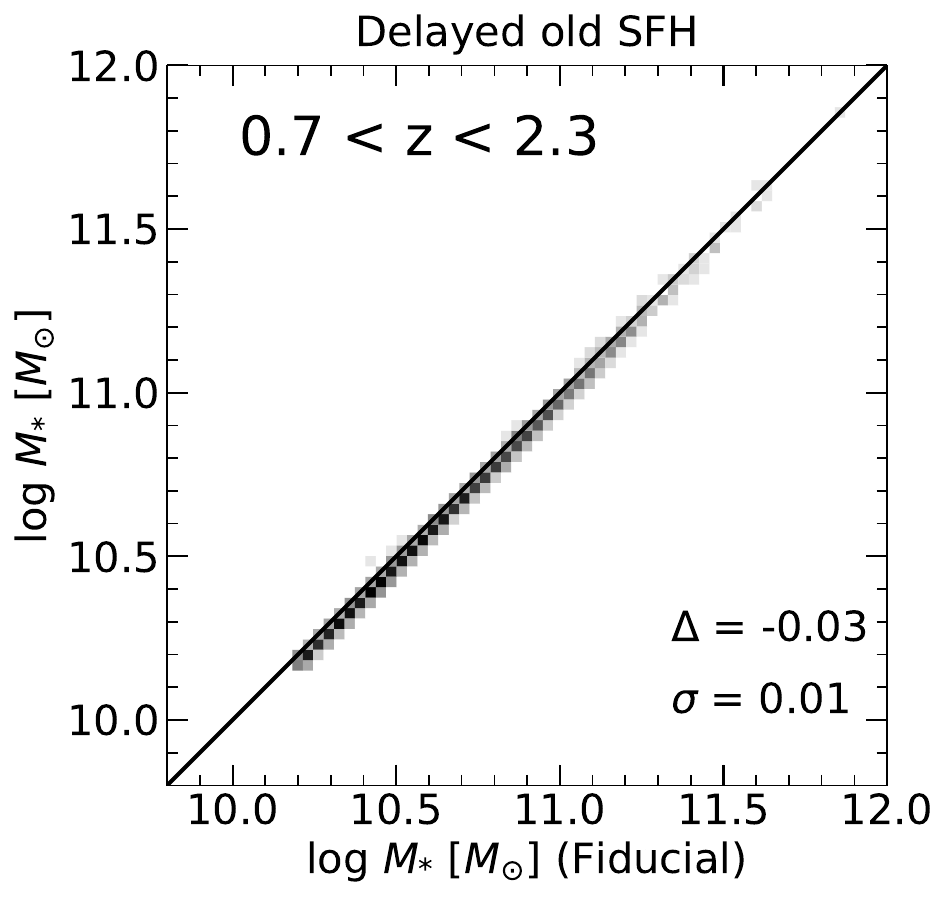}{0.3\textwidth}{}
          }
    \caption{Comparison of log stellar mass (log $M_{*}$) for various models to log $M_{*}$ from our fiducial model.  On average, the Calzetti and extra flexible models predict similar masses as the fiducial model while the flexible power law model predicts systematically higher masses compared to the fiducial model (top row).  The choice of metallicity does not result in systematic differences in mass; BPASS predicts lower masses compared to BC03 (middle row).  Masses are highly consistent between each SFH model (bottom row).  The scatter $\sigma$ is the standard deviation in log $M_* -$ log $M_{*\text{,fiducial}}$ while the offset $\Delta$ is the mean of log $M_*$ $-$ log $M_{*\text{,fiducial}}$.}
    \label{MassPlots}
\end{figure*}

So far we have focused on dust luminosity and therefore the SFRs. We now consider how the estimated stellar mass depends, in the relative sense, on the choice of model. The top row of Figure \ref{MassPlots} compares stellar masses from the Calzetti (left), extra flexible (middle), and flexible power law (right) models to stellar mass from the fiducial model.  In terms of the mean offset, the masses from both the Calzetti and extra flexible dust curves are consistent with fiducial masses for any individual galaxy.  The flexible power law model predicts higher stellar masses by $0.21$ dex on average compared to the fiducial model, which is significant considering that the typical error in the stellar mass (fiducial) for galaxies in our sample is $\sim 0.05$ dex. Similar result was found by \citet{LoFaro2017}, who report that masses based on power-law dust model are 0.15 dex higher than the ones from the Calzetti law. 

The middle row of Figure \ref{MassPlots} compares stellar masses from the free $Z_*$ BC03 (left), subsolar $Z_*$ BC03 (middle), and solar $Z_*$ BPASS (right) models to stellar mass from the solar $Z_*$ BC03 fiducial model.  Changing the metallicity treatment does not lead to significant systematic differences in mass.  However, BPASS predicts systematically lower masses compared to BC03, by $0.17$ dex on average; this is significant compared to the typical error in the stellar mass (fiducial) for galaxies in our sample, which is $\sim 0.05$ dex.  

The bottom row of Figure \ref{MassPlots} compares stellar masses from the flexible burst (left), younger flat bursts (middle), and delayed old SFH (right) models to stellar mass from the fiducial model.  Masses are quite similar between each model, with minimal scatter and offset.  While the flexible burst model has a noticeable one sided scatter, the bulk of the sample has essentially identical masses.  Notably, the delayed old SFH model also produces masses that are highly consistent with the fiducial model.  

\section{Catalog of SED fitting galaxy parameters}

To facilitate the use of our results in future studies we present a catalog of SED fitting galaxy parameters obtained using our fiducial model. Specifically, we publish our estimates of the stellar mass (log solar masses), SFR (log solar masses per year), FUV and $V$ band attenuation (magnitudes) and dust luminosity (log solar luminosities) together with their uncertainties. The catalog can be found online in both the journal and at \url{https://salims.pages.iu.edu/candels}. See Appendix \ref{sec:catalog} for some additional details and an example of rows and columns from the catalog. Included in the catalog are all galaxies for which we performed SED fitting with redshifts in the range $0.7 < z < 2.3$, with reduced $\chi_r^2 < 10$, and with differences in redshift between CANDELS and 3DHST of less than 0.4 ($|\Delta z| < 0.4$), amounting to 63,266 galaxies total. The CANDELS identifier (integer index used in all CANDELS catalogs) and CANDELS subfields (e.g., COSMOS, GOODS-S) are also included alongside the CANDELS redshifts used in the fitting (i.e., $z_{\text{best}}$). 

\section{Discussion}
\label{Section:Discussion}

In this section we discuss the most significant results outlined in the previous section and place them into the context of previous studies.  We also suggest some additional good practices for consideration in future studies.

\subsection{Dust attenuation}

Our results are generally consistent with previous findings that galaxy attenuation curves are diverse and not always well-described by a fixed law \citep{Kriek&Conroy2013DustLawsSlopeBumpCorrelation, Salmon2016NonUniversalDustLaw, Salim2018DustAttCurves, Theios2019BPASSdustAndStellarPopsAndMetallicity, Salim2020GalaxyDustAttReview}.  While a flexible dust treatment is essential for accurate SED fitting results, it is worth considering whether some constraint on the attenuation law could improve the agreement between $L_{\text{dust}}$ and $L_{\text{TIR}}$.  For instance, it has been shown that there is a relationship between the \textit{V}-band attenuation and the steepness of the attenuation curve \citep{Salim2020GalaxyDustAttReview}.  It has also been shown that a correlation exists between the power law slope and the strength of the UV bump \citep{Kriek&Conroy2013DustLawsSlopeBumpCorrelation}.  These correlations can be added as a constraint in the SED fitting, still allowing for flexibility but with an informed constraint on the parameter space; this saves computation time and could potentially reduce degeneracy in the SED fitting.  We tested both of these constraints separately but found that they offer no improvement in the agreement between $L_{\text{dust}}$ and $L_{\text{TIR}}$ for our sample.  However, such constraints could still potentially be useful if rest-frame UV photometry (essential for constraining the shape of the attenuation curve) is not available.  The $\delta - A_V$ and $\delta - \text{bump}$ relations could also serve as a physically-motivated model distribution in MCMC fitting frameworks such as Prospector, which uses a gridless MCMC-like approach to fit galaxy SEDs and includes the \citet{Kriek&Conroy2013DustLawsSlopeBumpCorrelation} slope-bump relation as a built-in distribution \citep[see][]{Leja2017ProspectorSEDCode}.  

We note that the shortest observed wavelengths included in the photometry are those spanned by the \textit{U} or \textit{u} bands (hereafter referred to collectively as \textit{U} bands).  The peak wavelength of the \textit{U} bands correspond roughly to 3500\AA{} which, at our lowest redshift ($z = 0.7$), corresponds to a rest-frame wavelength of $\sim 2000$\AA{} which is in the near-UV range.  One would ideally include both far-UV and near-UV coverage in the SED fitting since the UV color is sensitive to the dust.  It is only above roughly $z = 1$ that the \textit{U} bands cover the rest-frame far-UV; despite this, however, we find that the agreement between $L_{\text{dust}}$ and $L_{\text{TIR}}$ is actually better at $z \sim 1$ than at $z \sim 2$ regardless, possibly because of the higher S/N ratio of $z \sim 1$ photometry.  

\subsection{Stellar population synthesis models and metallicity}

We find that fixing the stellar metallicity is a preferred choice given the quality of the data we have in CANDELS and this type of sample; it may not be the best option in some other circumstances, e.g., for galaxies with old stellar populations where the dust has a lesser impact on the galaxy SED and the age and metallicity can be better constrained simultaneously \citep[see][]{Dorman2003AgeMetallicityConstraintsForOldPopsUsingMidUVColors}.  Our choice of solar metallicity may not be appropriate for galaxies outside the range of our sample, i.e., at lower masses (log $M_{\odot} < 10.2$) or higher redshifts ($z > 2.3$) for which the galaxies have yet to undergo significant enrichment.  It has also been found that the assumption of fixed metallicity may lead to systematic biases in the stellar mass, albeit for galaxies below our mass range of log $M_* < 10.2$ \citep{Mitchell2013reliabilityOfStellarMassFromBroadbandSEDFitting}.  

A more sophisticated approach than assuming a single fixed metallicity would be to impose a constraint on the metallicity according to some mass-dependent and perhaps redshift-dependent relation.  Such a technique was employed by \citet{Leja2019OlderMoreQuiescentUniverse} who, using the Prospector code sampled the mass-metallicity space according to a normal distribution with mean and sigma from the \citet{Gallazzi2005massmetallicitylocal} relation for local galaxies, clipping the distribution to the range $0.00021 < Z_* < 0.031$ ($-1.98 <$ log $(Z_* / Z_{\odot}) < 0.19$); this method effectively weights the models towards the local relation yet allows the fits to take on significantly lower values to account for possible evolution with redshift.

Another approach would be to fix the stellar metallicity in the fitting for a given mass and redshift based on an assumed mass-metallicity-redshift relation, effectively removing the metallicity as a free parameter (since redshift is fixed and the mass is tied to the SFH), which has been proposed by recent studies of the mass-metallicity relation in galaxies \citep[see][]{Thorne2022GalaxyMZRGama}.  Using VANDELS, \citet{Cullen2019VANDELSstellarmetallicitySFGs} found that the mass-metallicity relation for $z > 2.5$ star-forming galaxies is consistent in slope with the local relation, but is shifted in zero-point to lower values by $\sim 0.6$ dex.  A redshift-dependent mass-metallicity relation can therefore be introduced into the SED fitting by simply adopting the local relation and scaling it according to the galaxy redshift.  However, direct constraints on the stellar metallicity at intermediate redshifts (i.e., $0.3 < z < 2.5$) are severely limited due to the need for high-resolution rest-frame UV spectroscopy, which renders it difficult to determine the appropriate scaling with redshift.  One possible solution is to use the redshift evolution in the gas-phase metallicity-mass relation as a proxy for the scaling of the stellar mass-metallicity relation \citep[see e.g.,][]{Maiolino2008MassMetallicityRelationHighz, Wuyts2016evolutionOfMetallicityAtIntermediateRedshifts, Thorne2022GalaxyMZRGama}.  We note, however, that the resolution of the SPS model metallicity grid for some libraries may be too restrictive to implement a detailed redshift dependence, and our results may suggest that no redshift dependence is even needed, at least for $z \lesssim 2.3$, since we find, somewhat surprisingly, that solar metallicity models produce more reliable $L_{\text{dust}}$ compared to subsolar models at both $z \sim 1$ and $z \sim 2$. 

%We do not test mass-dependent metallicity models in this work given the limited mass range of our sample (log $M_{\odot} > 10.2$) but recommend their consideration for studies working across a wide range of masses. 

% maybe need to check if there is info on metallicity at intermediate redshifts

Whereas BC03 SPS models are known to compare favorably to other SPS models at low redshifts \citep[e.g.,][]{Conroy2010FSPSmodelingUncertaintiesPIII, Hansson2012colorComparisonugrizLocalSPSmodelComparison, Zibetti2013NIRspectraOfPostSBsSuggestLimitedImpactOfTPAGBsOnSEDs}, one might expect binary effects to be more relevant for galaxies at high redshift which have younger stellar populations and thus greater fractions of high-mass main sequence stars. It is thus interesting that BPASS should be disfavored by our results.  The effects of binary stars on the stellar emission may be obfuscated if the uncertainties in the UV photometry (in which regime the effects are most pronounced) are sufficiently large, while the treatment of other stellar evolution phases in the SPS models, such as TP-AGB stars or extreme horizontal branch stars, may have a stronger impact on the SED fitting overall \citep{Maraston2005StellarModelsTPAGB, vanderWel2006MassComparisonSPSandNearIRBiases, Conroy2009FSPSmodelingUncertaintiesPI, Walcher2011SEDfittingReview, Conroy2013SEDFittingReview, Zibetti2013NIRspectraOfPostSBsSuggestLimitedImpactOfTPAGBsOnSEDs, McGaugh2014ColorMtoLRelationsDiskGalaxies, Villaume2015galaxyNIRSEDsandAGBstars}.  It may be that BC03 models provide a more realistic treatment of certain aspects of non-binary stellar evolution compared to BPASS, which could explain why our results favor BC03 models.  However, we note that while BPASS is disfavored in terms of $L_{\text{dust}}$, we have no independent estimate of the stellar mass and so cannot rule out the possibility that BPASS produces more accurate masses.  We also cannot rule out the possibility that other properties determined from the fits, such as the strength of the ionizing continuum or specific spectral indices, may also be more realistic when using BPASS.  

% Mention BPASS applications, e.g. uv escape fraction or reionization studies

% The use of observationally-based stellar evolution tracks in BC03 may help explain why BPASS, which uses entirely theoretical stellar evolution prescriptions, is disfavored despite its inclusion of the effects of binary stars.  

% It is curious that BC03 with Z = 0.02 works best
% But BPASS with Z = 0.008 is very close.  BPASS may get things generally right but BC03 may get some aspects of stellar evolution better
% BPASS may give more realistic ages and masses even if Ldust doesnt match quite as well, but again the difference is slight between best BPASS and best BC03
% BPASS may be more realistic overall and the best choice of metallicity matching what we expect empirically may support this
% BPASS already has been used for high-z reionization studies, cite these maybe
% Does BC03 have an advantage over BPASS since BPASS is 100 percent theoretical??  Difficult to really argue since the binary evolution may be more significant, but both could matter

\subsection{Star Formation History}

The results of our tests on SFH suggest that minimizing the variability of the very recent ($\lesssim 100$ Myr) SFH produces the most reliable $L_{\text{dust}}$.  We find that this is attributable to age-dust degeneracy; when allowing for sharp drops in the recent SFH, models with less dust and a recent drop in the SFR will produce similar colors to a model with constant or rising recent SFR and more dust.  The fiducial flat burst model resolves this issue by forcing the recent ($< 1$ Gyr) SFH to either increase or remain flat, effectively avoiding the age-dust degeneracy.  Our findings echo those of \citet{Wuyts2011GalaxyStarFormationOF44}, who find, using single $\tau$ SFH models with variable age, that allowing for very small e-folding times ($\lesssim 300$ Myr) results in less accurate SFRs.  However, \citet{CurtisLake2021BiasInMSScatterAtZof5UsingMockGalaxies} find that for relatively low mass galaxies ($8 \lesssim$ log $M_* \lesssim 10.5$) at very high redshifts ($z \sim 5$), broadband photometry becomes sensitive to fluctuations in the SFR on shorter timescales ($\sim 10$ Myr), so in such a regime it may be necessary to allow for more stochasticity in the recent SFH to obtain accurate physical parameters.

We also note that the so-called `post-starburst' or `E+A' galaxies feature recent ($< 1$ Gyr) quenching in their SFH \citep[e.g.,][]{French2021PostSBE+AReview}.  A recent quenching is not allowed by our fiducial SFH.  It may be necessary to allow for more flexibility in the recent SFH to obtain accurate physical parameters for post-starbursts specifically \citep[see][]{Suess2022PostStarburstSFHrecovery}.  However, post-starbursts are relatively rare \citep[see][]{French2021PostSBE+AReview} and they should not significantly affect the results from SED fitting of statistical samples.  Furthermore, post-starbursts are sometimes selected based on the ratio of $H\alpha$ to UV SFR, so it is unclear whether $L_{\text{dust}}$ would even be significantly affected as the SED fitting may not be sensitive to changes in the SFR on such short timescales, i.e., on the order of $10$ Myr \citep{FloresVelazquez2021UVHaTimescalesFIREsims, French2021PostSBE+AReview}.  

Our findings regarding the use of delayed versus exponential parameterizations for the old SFH are consistent with past studies which find that the details of the old SFH are effectively unconstrained in broadband SED fitting, though parametric forms where the age of the old population is allowed to vary freely are susceptible to underestimating stellar masses due to outshining \citep{Salim2016GSWLC, Carnall2019parametricSFHs, Leja2019nonparametricSFHs, Lower2020StellarMassSEDFittingSFHBias}.  So long as the outshining bias is accounted for (which we accomplish by fixing the age of the old population to near the maximum possible time), the systematic uncertainties in $M_{*}$ and $L_{\text{dust}}$ are dominated by the dust, metallicity, stellar population models, and/or recent SFH, rather than the ancient SFH.  
% maybe we can say that when mass is handled fine then ages should be fine too

\subsection{Published catalogs of physical properties from SED fitting}

We note that value-added catalogs of physical properties determined via UV-NIR SED fitting are already available for thousands of galaxies at a wide range of redshifts in certain CANDELS fields.  One example is the \citet{Santini2015GOODSSMassCatalog} catalog which includes physical properties for thousands of GOODS-S and UDS galaxies compiled from various groups who used different SED fitting codes and model assumptions; notably, some groups allow the stellar metallicity to vary while others keep it fixed, some groups use SPS models other than that of BC03, some groups use single-component SFHs, and a fixed \citet{Calzetti2000SFGDust} dust law is assumed in all but one instance (in which case the group also allowed for the SMC dust law).  Another public catalog is the \citet{Barro2019CANDELSGOODSNpropertiesCatalog} catalog for GOODS-N which includes stellar masses and SFRs based on both UV-NIR SED fitting as well as combined UV+IR SFRs. Their SED fitting method is based on that of \citet{Wuyts2011GalaxyStarFormationOF44}.  They find the different estimates of SFR to be generally consistent for galaxies with both UV and IR photometry.  While they hold the stellar metallicity fixed and adopt BC03 SPS models, they also assume a fixed \citet{Calzetti2000SFGDust} dust law and use single-component parametric SFHs with variable age.  For users of these catalogs we caution that the assumption of single component SFHs may lead to underestimated stellar masses \citep[e.g.,][]{Lower2020StellarMassSEDFittingSFHBias} and the assumption of a fixed dust law or variable stellar metallicity may result in biased SFRs for some galaxies as we have demonstrated \citep[see also][]{Salmon2016NonUniversalDustLaw, Salim2018DustAttCurves}.   

\begin{figure}
    \centering
    \gridline{
          \fig{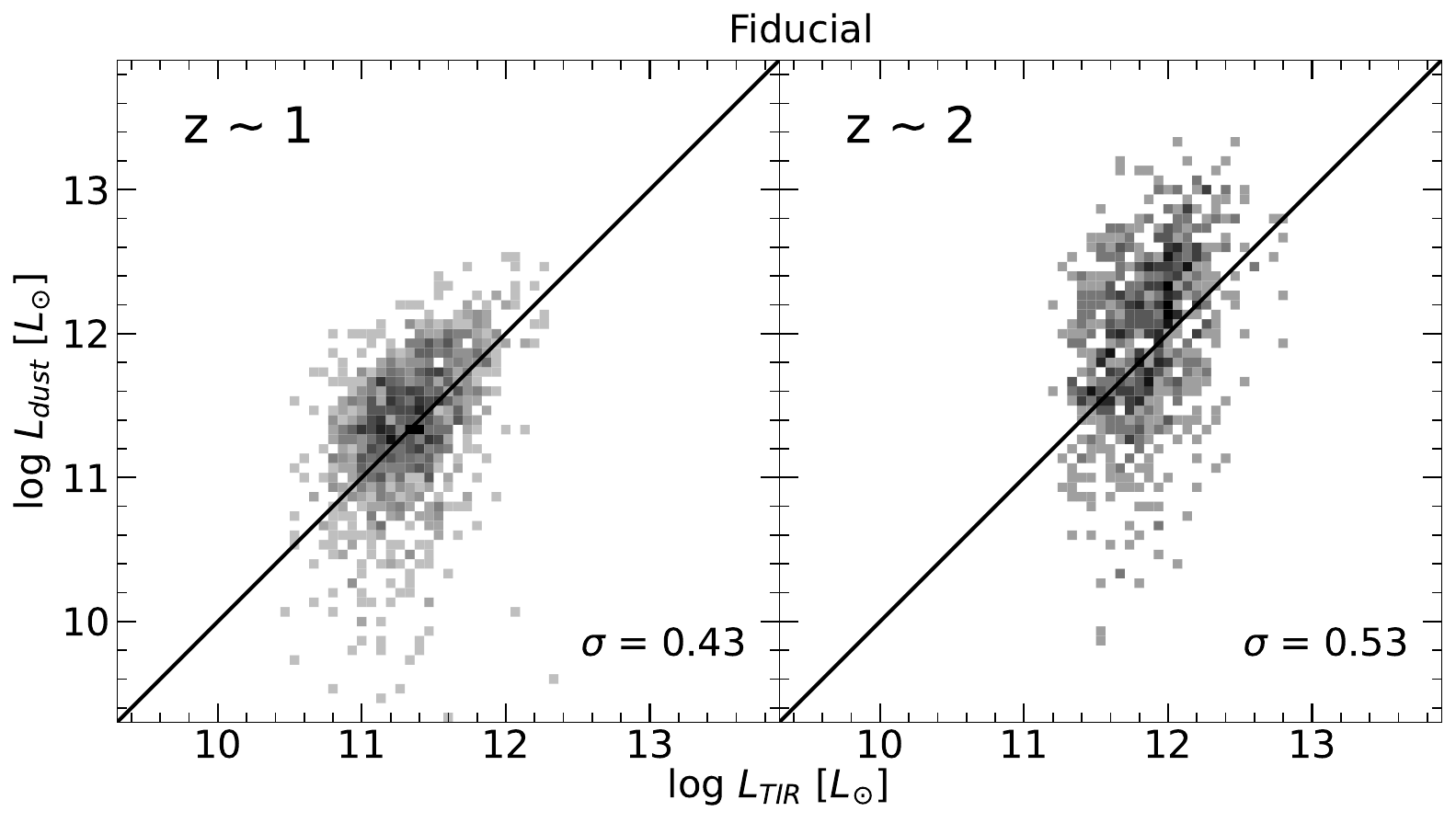}{0.45\textwidth}{}
    }
    \vspace*{-\baselineskip}
    \gridline{
          \fig{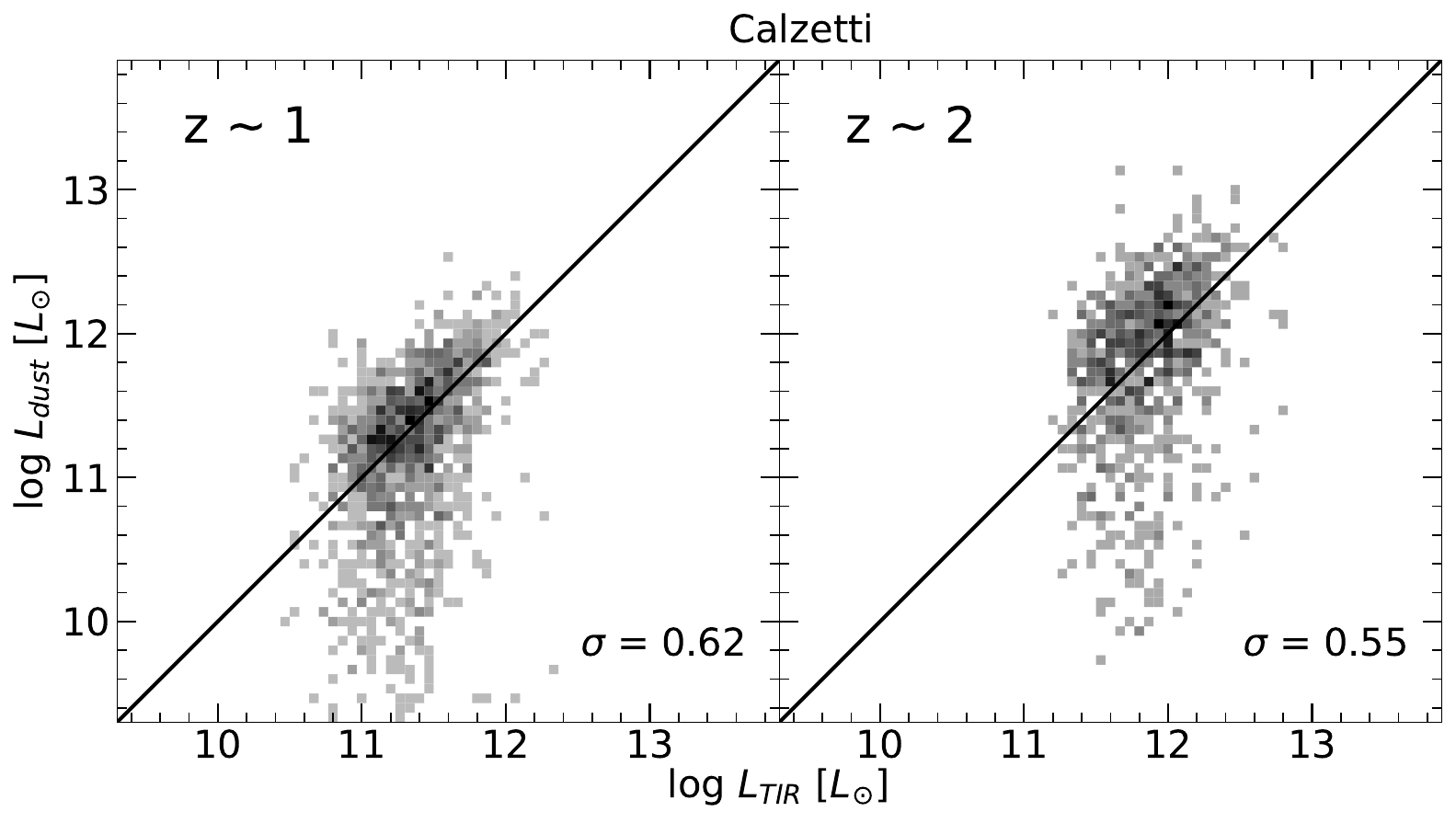}{0.45\textwidth}{}
    }
    \vspace*{-\baselineskip}
    \gridline{
          \fig{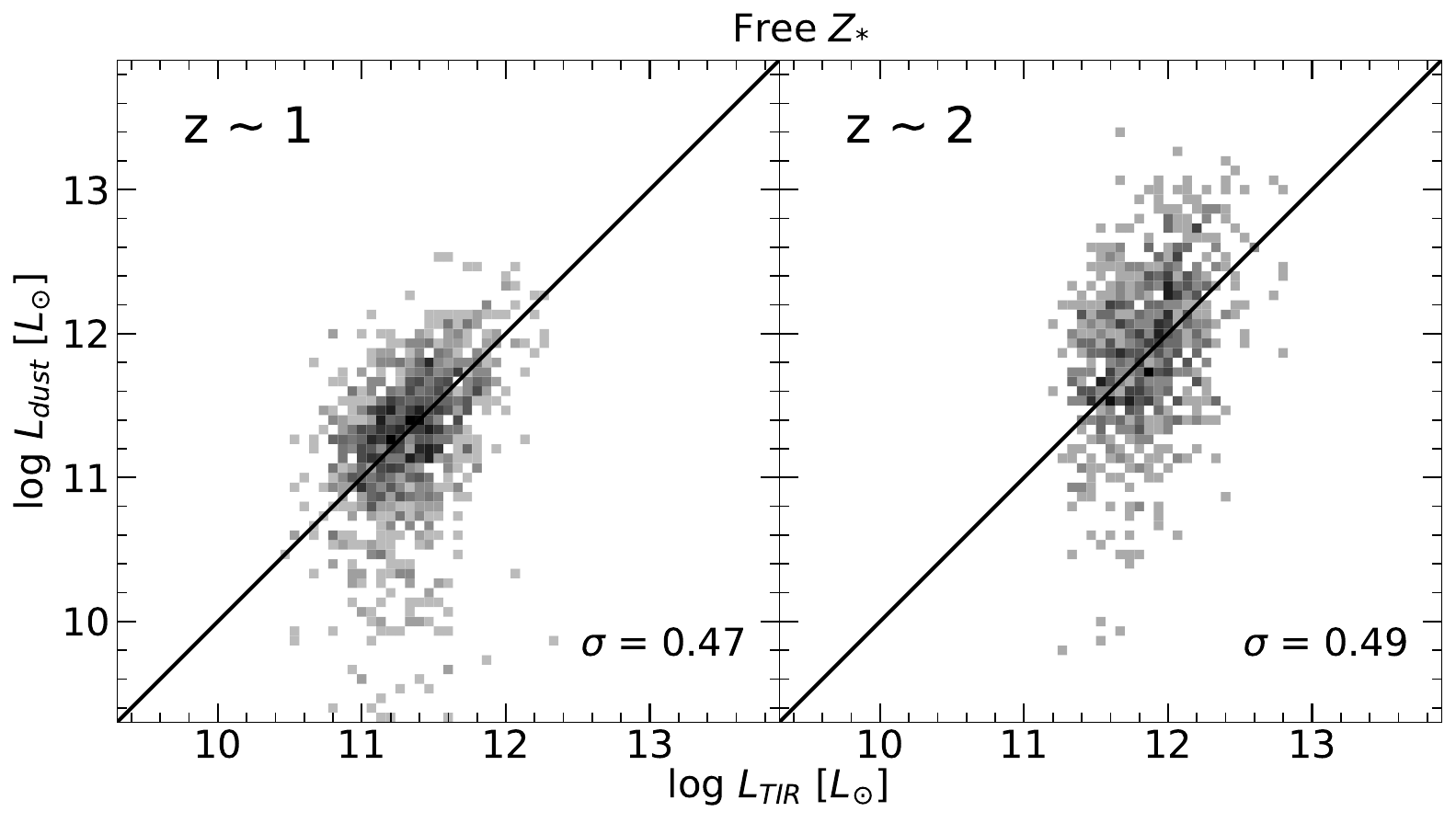}{0.45\textwidth}{}
    }
    \vspace*{-\baselineskip}
    \gridline{
          \fig{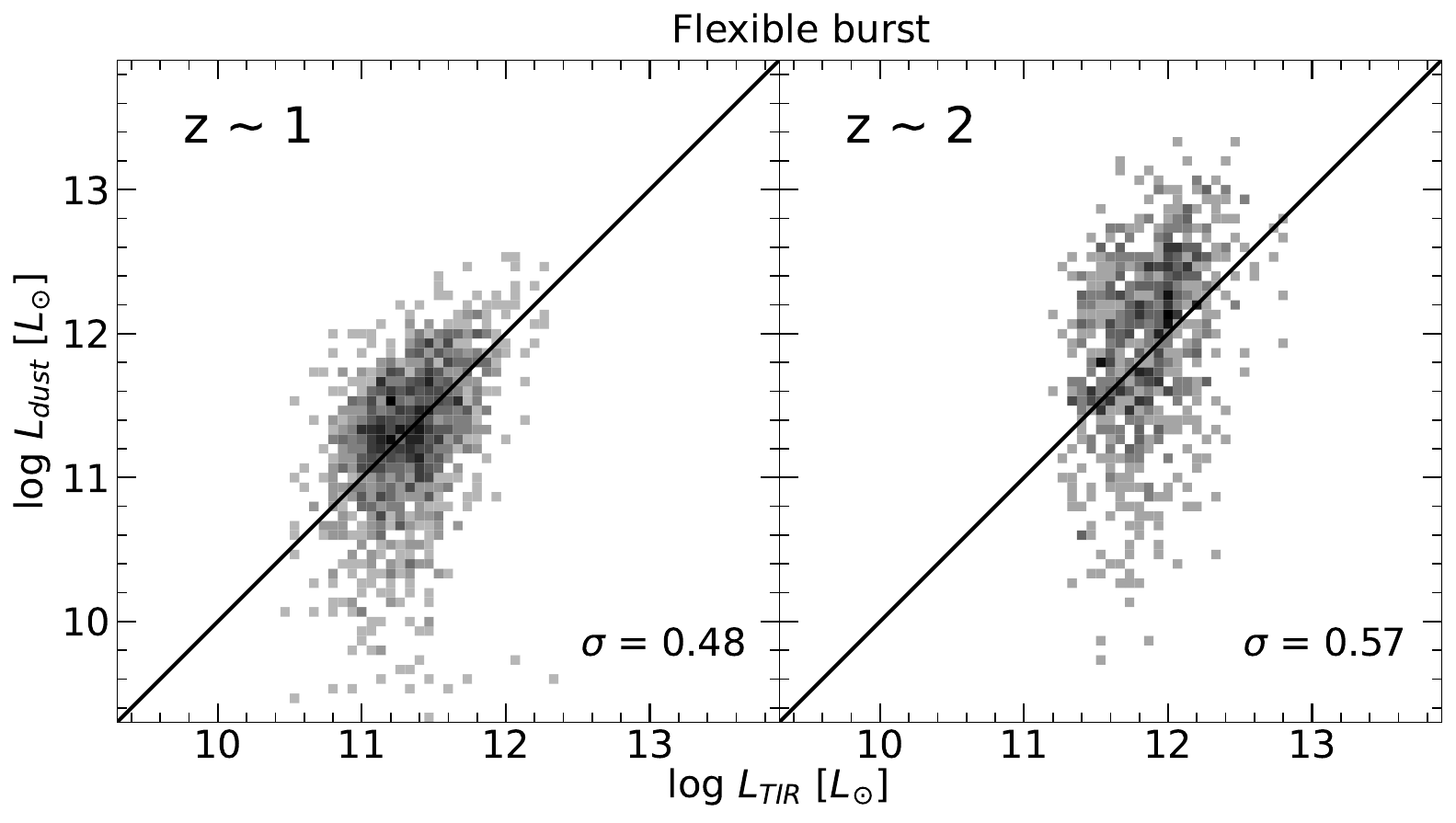}{0.45\textwidth}{}
    }
    \vspace*{-\baselineskip}
    \caption{Comparison of $L_{\text{dust}}$ from UV-optical-NIR SED fitting to $L_{\text{TIR}}$ for different models using photometry re-sampled with artificially increased errors to mimic photometry with 2 mag shallower limiting magnitudes.  The relative performance of the different models is similar to the original results.}
    \label{PoorRunPlots}
\end{figure}

\subsection{Applicability to other datasets}

It is worth considering whether the results obtained in this study, which features a sample of high-mass galaxies with abundant UV-optical photometry from CANDELS, is applicable in the hypothetical case of shallower observational data.  To test this, we perform a set of mock SED fitting where for each galaxy we artificially increase the photometric errors by a factor of 6.3 (thus mimicking 2 mag brighter limiting magnitudes), then resample the fluxes using the larger errors.  We then fit a selection of our models to determine whether their relative performance is affected by the decrease in signal-to-noise.  Figure \ref{PoorRunPlots} shows the results of this mock fitting for fiducial, free $Z_*$, Calzetti dust, and flexible burst models.  We find the relative performance of each model is consistent with our main results; fiducial performs best overall, though free $Z_*$ outperforms fiducial for $z \sim 2$ galaxies.  The consistency in results despite the larger errors suggests that our results are applicable even to fields with significantly shallower photometry than that typical of CANDELS. 

\subsection{Choice of SED fitting code}

We do not explore the impact of different SED fitting codes in this work.  However, different codes come pre-packaged with different model libraries and the underlying fitting `machinery' (i.e., the inference method used to estimate properties from fitted models) may also differ.  These differences may result in systematic uncertainties and, especially in the case of SFRs and dust attenuation (e.g., $A_V$), systematic offsets which depend on the choice of SED fitting code \citep[][]{Pacifici2023CriticalAssessmentOFSEDFittingTechniques}.  We refer to \citet{Pacifici2023CriticalAssessmentOFSEDFittingTechniques} for a discussion of methodological best practices which can be used to mitigate the impact of code-dependent systematics.

\section{Conclusions}
\label{Section:Conclusions}

In this work, we are able to identify some good practices for SED fitting at $z \gtrsim 1$ to improve constraints on SFR and reduce the effects of age-dust-metallicity degeneracy.  Using a sample of star-forming galaxies with abundant UV-optical-NIR photometry from CANDELS and coverage in the IR with Spitzer, we evaluate which SED fitting models produce the best agreement (in terms of scatter) between $L_{\text{dust}}$ from the UV-optical-NIR fitting and $L_{\text{TIR}}$ inferred from the $24\mu$m photometry.  Our use of dust (total IR) luminosity as a point of comparison, rather than SFR directly, avoids systematics due to heating of dust by old stars and the choice of an SFR timescale.  

We find generally that for SED fitting there exists a `sweet spot' between assuming a fixed model on the one hand, and having too much freedom in the models on the other hand.  Going too far in either direction can result in the introduction of noise or systematics.  Whereas exploring different flavors of SED fitting allowed us to identify better choices, we have yet to arrive at a precise match between $L_{\text{dust}}$ and $L_{\text{TIR}}$ in terms of zero point differences and the skew (i.e., linearity).  Our main conclusions are as follows:

\begin{enumerate}

\item Allowing for dust attenuation curve slopes to be flexible, and including the slopes that would be considered rather steep (SMC-like) , is essential for getting unbiased $L_{\text{dust}}$ (and thus presumably SFRs). On the other hand, assuming a fixed shallow curve can lead to a number of galaxies with severely underestimated dust luminosities.

\item The flexible dust attenuation model of \citet{Noll2009CIGALE}, which takes the Calzetti curve and allows its steepness to vary, produces better results than when the form of the underlying curve is assumed to be a pure power law. 

\item Assuming a fixed stellar metallicity, in particular the solar metallicity, produces better results than allowing the metallicity to vary.
  
\item BC03 SPS models produce $L_{\text{dust}}$ estimates that are in tighter agreement with observations than the ones produced by BPASS models. 

\item A SFH model in which the recent burst of SF has constant SFR over at least the past 100 Myr produces better results than allowing for a more recent burst, or allowing the burst to decline.

\item Whether one assumes an exponentially declining SFH for the first (older) component or the so called delayed SFH makes essentially no difference, presumably because the shape of the old SFH is poorly constrained by the broadband photometry.

\item Stellar masses between different models generally agree to within few hundredths of a dex. The exception is when the attenuation curve slopes are based on power laws, which results in 0.21 dex higher masses than our fiducial dust model, and when using BPASS models, which results in 0.17 dex lower masses than with BC03 models. 

\item The relative performance of different models remains similar when shallower photometry is available. 

\end{enumerate}

We also make publicly available our estimates of the stellar mass and SFR, among other parameters.

This work was supported through NASA award 80NSSC20K0440.

\appendix
\section{Catalog Format}\label{sec:catalog}

In this section we illustrate the format of the published catalog of galaxy physical parameters associated with this work. The data and sample selection for the catalog are described briefly in Section 5. Table \ref{tab:catalog} shows a small selection of galaxies and their properties from the catalog. The catalog is accessible from both the journal and from the website \url{https://salims.pages.iu.edu/candels}. 

%
%\begin{rotatetable}
\begin{splitdeluxetable*}{ccccccccBccccccc}
\tablecaption{A selection of rows from the published catalog of SED fitting parameters. Parameters are estimated from SED fitting with fiducial models. \label{tab:catalog}}
%\tabletypesize{\tiny}
\tablehead{\colhead{Seq} & \colhead{Field} & \colhead{RA} & \colhead{DEC} & \colhead{logMstar} & \colhead{e\_logMstar} & \colhead{logSFR} & \colhead{e\_logSFR} & \colhead{AFUV} & \colhead{e\_AFUV} & \colhead{AV} & \colhead{e\_AV} & \colhead{logLdust} & \colhead{e\_logLdust} & \colhead{$z_{\text{best}}$}\\ \colhead{ } & \colhead{ } & \colhead{deg} & \colhead{deg} & \colhead{dex($M_{\odot}$)} & \colhead{dex($M_{\odot}$)} & \colhead{dex($M_{\odot}$ / yr)} & \colhead{dex($M_{\odot}$ / yr)} & \colhead{$\mathrm{mag}$} & \colhead{$\mathrm{mag}$} & \colhead{$\mathrm{mag}$} & \colhead{$\mathrm{mag}$} & \colhead{dex($L_{\odot}$)} & \colhead{dex($L_{\odot}$)} & \colhead{ }}
\startdata
22 & GOODS-S & 53.09084 & -27.95640 & 9.379 & 0.157 & 0.616 & 0.195 & 1.746 & 0.467 & 0.820 & 0.257 & 10.529 & 0.156 & 0.905 \\
36 & GOODS-S & 53.10218 & -27.95356 & 9.168 & 0.134 & 0.054 & 0.192 & 1.622 & 0.500 & 0.494 & 0.243 & 9.963 & 0.189 & 0.761 \\
51 & GOODS-S & 53.07999 & -27.95205 & 10.038 & 0.213 & 0.429 & 0.530 & 3.133 & 1.437 & 0.804 & 0.494 & 10.636 & 0.278 & 0.759 \\
64 & GOODS-S & 53.07955 & -27.95027 & 9.498 & 0.156 & 0.386 & 0.201 & 0.957 & 0.515 & 0.256 & 0.198 & 10.188 & 0.252 & 0.823 \\
78 & GOODS-S & 53.09077 & -27.94879 & 8.813 & 0.168 & -0.183 & 0.193 & 0.976 & 0.523 & 0.214 & 0.189 & 9.597 & 0.247 & 0.911 \\
90 & GOODS-S & 53.07146 & -27.94784 & 9.474 & 0.158 & 0.319 & 0.244 & 1.489 & 0.616 & 0.507 & 0.295 & 10.246 & 0.227 & 0.922 \\
111 & GOODS-S & 53.09655 & -27.94672 & 9.524 & 0.119 & 0.316 & 0.155 & 0.791 & 0.435 & 0.236 & 0.181 & 10.060 & 0.243 & 0.733 \\
115 & GOODS-S & 53.09890 & -27.94580 & 8.876 & 0.112 & -0.370 & 0.132 & 0.783 & 0.380 & 0.198 & 0.161 & 9.353 & 0.224 & 0.806 \\
\enddata
\tablecomments{Table \ref{tab:catalog} is published in its entirety in the machine-readable format. 
A portion is shown here for guidance regarding its form and content.}
\end{splitdeluxetable*}
%\end{rotatetable}

\bibliography{Bibliography}
\bibliographystyle{aasjournal}

\end{document}